\PassOptionsToPackage{unicode}{hyperref}
\PassOptionsToPackage{hyphens}{url}
\PassOptionsToPackage{dvipsnames,svgnames,x11names}{xcolor}
\documentclass[
  11pt,
]{article}
\usepackage{xcolor}
\usepackage[margin=2.5cm]{geometry}
\usepackage{amsmath,amssymb}
\usepackage{iftex}
\ifPDFTeX
  \usepackage[T1]{fontenc}
  \usepackage[utf8]{inputenc}
  \usepackage{textcomp} 
\else 
  \usepackage{unicode-math} 
  \defaultfontfeatures{Scale=MatchLowercase}
  \defaultfontfeatures[\rmfamily]{Ligatures=TeX,Scale=1}
\fi
\usepackage{lmodern}
\ifPDFTeX\else
\fi
\IfFileExists{upquote.sty}{\usepackage{upquote}}{}
\IfFileExists{microtype.sty}{
  \usepackage[]{microtype}
  \UseMicrotypeSet[protrusion]{basicmath} 
}{}
\makeatletter
\@ifundefined{KOMAClassName}{
  \IfFileExists{parskip.sty}{%
    \usepackage{parskip}
  }{
    \setlength{\parindent}{0pt}
    \setlength{\parskip}{6pt plus 2pt minus 1pt}}
}{
  \KOMAoptions{parskip=half}}
\makeatother
\usepackage{longtable,booktabs,array}
\usepackage{caption}
\usepackage{calc} 
\usepackage{etoolbox}
\makeatletter
\patchcmd\longtable{\par}{\if@noskipsec\mbox{}\fi\par}{}{}
\makeatother
\IfFileExists{footnotehyper.sty}{\usepackage{footnotehyper}}{\usepackage{footnote}}
\makesavenoteenv{longtable}
\NewDocumentCommand\citeproctext{}{}

\makeatletter
 \let\@cite@ofmt\@firstofone
 \def\@biblabel#1{}
 \def\@cite#1#2{{#1\if@tempswa , #2\fi}}
\makeatother
\newlength{\cslhangindent}
\newlength{\csllabelwidth}
\newenvironment{CSLReferences}[2] 
 {\begin{list}{}{%
  \setlength{\itemindent}{0pt}
  \setlength{\leftmargin}{0pt}
  \setlength{\parsep}{0pt}
  \ifodd #1
   \setlength{\leftmargin}{\cslhangindent}
   \setlength{\itemindent}{-1\cslhangindent}
  \fi
  \setlength{\itemsep}{#2\baselineskip}}}
 {\end{list}}
\usepackage{calc}

\newcommand{\CSLLeftMargin}[1]{\parbox[t]{\csllabelwidth}{\strut#1\strut}}
\newcommand{\CSLRightInline}[1]{\parbox[t]{\linewidth - \csllabelwidth}{\strut#1\strut}}

\providecommand{\tightlist}{%
  \setlength{\itemsep}{0pt}\setlength{\parskip}{0pt}}
\usepackage{amsmath}
\usepackage{amssymb}
\DeclareUnicodeCharacter{2212}{\ensuremath{-}}
\usepackage{titlesec}
\titleformat{\part}[block]{\Large\bfseries\centering}{}{0pt}{}
\titlespacing*{\part}{0pt}{2ex}{1.5ex}
\usepackage{xurl}
\usepackage{etoolbox}
\AtBeginEnvironment{longtable}{\small}
\AtBeginDocument{\setlength{\csllabelwidth}{3.6em}}
\AtBeginEnvironment{verbatim}{\footnotesize}
\usepackage{fontawesome5}
\newcommand{\topicbme}{\faBalanceScale}
\newcommand{\topicphylo}{$\pitchfork$}
\newcommand{\topicpoly}{\tikz[baseline=-0.55ex,scale=0.13,line width=0.6pt]\draw (30:1)--(90:1)--(150:1)--(210:1)--(270:1)--(330:1)--cycle;}
\newcommand{\topicsoftware}{\faCog}
\usepackage{tikz}
\usepackage{subcaption}
\usetikzlibrary{calc}
\usepackage{algorithm}
\usepackage{algpseudocode}
\algnewcommand{\IIf}[2]{\State\algorithmicif\ #1\ \algorithmicthen\ #2}
\algnewcommand{\Continue}{\textbf{continue}}
\algnewcommand{\Break}{\textbf{break}}
\algrenewcommand{\algorithmiccomment}[1]{\hfill$\triangleright$\ #1}
\newcommand{\sv}[1]{\textsf{#1}}
\usepackage{bookmark}
\IfFileExists{xurl.sty}{\usepackage{xurl}}{} 
\makeatletter
\@ifundefined{xmpquote}{}{}
\makeatother
\hypersetup{
  pdftitle={A Compact Weight-Space Integer Programming Formulation for Balanced Minimum Evolution},
  pdfauthor={Leonid Chindelevitch\^{}\{1,*\}; Matthew Penn\^{}\{2\}; Samir Bhatt\^{}\{1,2,*\}},
  colorlinks=true,
  linkcolor={blue},
  filecolor={Maroon},
  citecolor={Blue},
  urlcolor={blue},
  pdfcreator={LaTeX via pandoc}}

\title{A Compact Weight-Space Integer Programming Formulation for Balanced Minimum Evolution}
\author{Leonid Chindelevitch\(^{1,*}\) \and Matthew Penn\(^{2}\) \and Samir Bhatt\(^{1,2,*}\)}
\date{}

\begin{document}
\maketitle

\begin{center}\small
$^{1}$ MRC Centre for Global Infectious Disease Analysis, School of Public Health, Imperial
College London, London, United Kingdom\\
$^{2}$ Section of Health Data Science and AI, Department of Public Health, University of
Copenhagen, Copenhagen, Denmark\\[1ex]
Emails: \texttt{lchindel@ic.ac.uk}, \texttt{mattpenn114@gmail.com}, \texttt{samir.bhatt@sund.ku.dk}\\
$^{*}$ Co-corresponding authors
\end{center}

\section{Abstract}\label{abstract}

Balanced Minimum Evolution (BME) reconstructs an unrooted binary phylogenetic tree from a distance matrix \(D\) by minimizing the Pauplin length \(\sum_{i<j}D_{ij}w_{ij}\), which is linear in the weights \(w_{ij}=2^{-\tau_{ij}}\) (\(\tau_{ij}\) the number of edges on the \(i\)--\(j\) path). We study the formulation of BME in these \(\binom n2\) weight variables.

\emph{Theoretical results.} (i) For a fixed topology \(T\), the Kraft equalities, the split-cut equalities \(\{W[S]=\tfrac12\}\) of \(T\), and the single convex constraint \(\sum w_{ij}\log_2 w_{ij}\le\tfrac32-n\) admit exactly one point, the tree's weight vector; the proof is a one-line application of the log-sum inequality. (ii) For every tree and every bipartition, the cut value \(W_T[S]\) lies in \(\{\tfrac12\}\cup\{\tfrac34\}\cup
[\tfrac78,\infty)\); the value \(\tfrac34\) is attained exactly by the \(2(n-3)\) bipartitions created by one nearest-neighbour interchange; and the number of bipartitions with \(W_T[S]\le1\) is \(2n^2-10n+15\) for every binary topology. (iii) The crossing graph of nontrivial bipartitions has an explicit edge clique cover with \(O(n^4)\) cliques, and every edge clique cover needs \(\Omega(n^2)\) cliques. (iv) The necessary conditions of Catanzaro, Pesenti and Wolsey for a path-length matrix, without the four-point condition, are sufficient for \(n\le11\) and first fail at \(n=12\) (a computer-assisted result). Catanzaro \emph{et al.} published an \(n=12\) counterexample; we add the threshold, the uniqueness of the counterexample up to relabelling, and a construction that explains it.

\emph{Computational results.} A compact branch-and-bound solver built on these results, which commits and forbids splits and imposes the convex constraint exactly, certifies optimality on benchmark instances with up to \(n=20\) taxa within an hour. On the same machine it is faster than the distance-indexed model of Catanzaro \emph{et al.} on two of these instances and slower on the others; that model, whose root bound is tighter, certifies 13 of 14 benchmark instances with \(n\le30\) within an hour in every seed. Ablations show that no particular cut family determines its root bound. Its advantage comes from lifting to integer level indicators, which makes the whole entropy equality satisfied by trees linear and lets the MIP solver's general-purpose cutting planes tighten the root; separating the four-point condition during the search then makes its certificates correct and fast. In \(w\) only half of this entropy equality is convex, and the compact model offers no integer structure of
this kind.

Although the compact solver does not improve on the computational state of the art overall, we explain why, and we discuss the open problems that our results raise.

\section{1. Introduction}\label{introduction}

Balanced minimum evolution is a distance-based criterion for phylogenetic reconstruction. Pauplin {[}1{]} showed that the length of a tree fitted to a distance matrix under the balanced weighting is a linear function of the matrix, with coefficients determined by the tree's topology alone. The criterion is statistically consistent {[}2{]}. It underlies the FastME software {[}3{]}, and Neighbor-Joining (NJ) {[}4{]} is a greedy algorithm for it {[}5{]}. NJ does not always return the BME-optimal tree {[}6{]}, so finding the optimum is a separate problem.

On the theoretical side, the Balanced Minimum Evolution Problem (BMEP) is NP-hard and has no polynomial-time \(c^{\,n}\)-approximation for some \(c>1\) unless P\(=\)NP {[}7{]}; the variant with a fixed tree shape, which only assigns the taxa to the leaves of a given tree, is also NP-hard and also hard to approximate {[}8{]}. The convex hull of the weight vectors of all trees, the BME polytope, has been studied through its facets {[}9{]}, {[}10{]} and its faces {[}6{]}, {[}11{]}, and Catanzaro \emph{et al.} characterize the integer path-length matrices of binary trees by linear and disjunctive conditions {[}12{]}, {[}13{]}. A survey of the area is given in {[}14{]}.

On the computational side, exact methods are integer programs solved by branch-and-bound or branch-and-cut {[}15{]}, {[}16{]}, {[}17{]}. The current best, the distance-indexed branch-and-cut of Catanzaro \emph{et al.} {[}13{]}, is reported to certify instances of up to about 32 taxa within an hour {[}18{]}. These formulations all work with integer path lengths and indicators of their values, which need \(\Theta(n^3)\) variables.

The objective, however, is linear in the weights \(w_{ij}=2^{-\tau_{ij}}\), of which there are only \(\binom n2\). This paper asks what a formulation in these weights can and cannot do.

Our starting point is a pinning result (Theorem 1). Let us fix a tree \(T\), and consider the following constraints on nonnegative vectors \(w\): every bipartition of the leaves induced by an edge of \(T\) (a split of \(T\)) has a cut value, the total weight of the pairs of leaves it separates, of \(\tfrac12\); and an entropy of \(w\) satisfies a lower bound valid for all trees. The weight vector of \(T\) is the unique point that satisfies all these constraints simultaneously. Without the entropy bound, the points satisfying the split constraints form a polytope that, for \(n\ge5\), is far from a single point. So in \(w\) a tree is described exactly by \(2n-3\) linear equalities and one convex inequality, without integrality. This motivates the rest of the paper: committing splits as equalities is how a search in \(w\) narrows down to a tree, the entropy bound is what turns a full set of committed splits into a single point, and the solver of Part II branches by committing or
forbidding splits for this reason. Table 1 lists our main results.

\textbf{Table 1. The main results of this paper and their relationship to previous work.}

{\def\LTcaptype{none} 
\begin{longtable}[]{@{}ll@{}}
\toprule\noalign{}
result & relationship to previous work \\
\midrule\noalign{}
\endhead
\bottomrule\noalign{}
\endlastfoot
Thm 1: split equalities and one convex constraint pin a fixed tree & new; elementary \\
Thm 2: the cut-value ladder; \(W=\tfrac34\) exactly at one NNI & new; refines the floor \(W\ge\tfrac12\) \\
Thm 3: \(\lvert\{S:W_T[S]\le1\}\rvert=2n^2-10n+15\) for every tree & new \\
Prop 4: \(O(n^4)\) and \(\Omega(n^2)\) bounds on crossing-graph clique covers & new \\
Thm 5: the conditions without four-point suffice if and only if \(n\le11\) & new; computer-assisted \\
the unique \(n=12\) counterexample and its construction & new; existence due to {[}13{]} \\
Prop G.1: four-point on the quartets containing one leaf implies all & short proof of {[}13{]}, thm. 2 \\
a linearity trade-off between \(w\) and \(\tau\) & observation \\
the compact solver against the distance-indexed solver & computational \\
\end{longtable}
}

The rest of this paper is organised as follows. Part I contains the theoretical results. Section 2 sets out definitions and the two formulations. Section 3 proves the pinning theorem, the cut-value ladder and the invariant count, and Section 4 reviews the known facets of the BME polytope. Section 5 gives the polynomial clique cover and encodes splits by membership variables. Section 6 compares linearity in \(w\) and in \(\tau\) and settles the sufficiency question. Part II contains the computational results: Section 7 describes the experimental setup and Section 8 the experiments. Section 9 discusses open problems. The appendices contain the proof of an inequality used in Section 3.2 (Appendix A), the proofs of Theorems 2 and 3 (Appendix B), the enumeration behind Theorem 5 (Appendix C), the tables of results behind Section 8 (Appendix D), the checks of numerical reliability (Appendix E), the compact solver's branch and bound in pseudocode (Appendix F), and a short proof that the
four-point and triangle conditions on the quartets and triples that contain one leaf imply them everywhere (Appendix G), and the lower bound of Proposition 4 with a smaller clique cover (Appendix H).

\part*{Part I: Theoretical results}

\section{2. Preliminaries}\label{preliminaries}

In this section we fix the notation for trees, splits and tree metrics (Section 2.1), state the problem (Section 2.2), introduce the weight space in which we work (Section 2.3), and set out the two integer programming models that the paper compares (Section 2.4).

\subsection{2.1 Trees, splits and tree metrics}\label{trees-splits-and-tree-metrics}

A \textbf{phylogenetic tree} on \([n]=\{1,\dots,n\}\) is a tree whose leaves are in bijection with \([n]\) and none of whose internal vertices has degree \(2\). All trees are unrooted. A tree is \textbf{binary} if every internal vertex has degree \(3\); it then has \(2n-3\) edges (\(n\) pendant, \(n-3\) internal) and \(n-2\) internal vertices. Contracting internal edges gives non-binary trees; the extreme case is the \textbf{star}, with one internal vertex. A \textbf{cherry} is a pair of leaves adjacent to the same internal vertex. The \textbf{topology} of a tree is the tree as a graph with labelled leaves, without edge lengths; we identify a tree with its topology.

A \textbf{bipartition} is an unordered pair \(\{S,S^c\}\) with both blocks nonempty, where \(S^c=[n]\setminus S\) is the complement of \(S\). We name it by its smaller block, and when both blocks have \(n/2\) leaves, by the block not containing leaf \(1\). It is \textbf{trivial} if one block is a single leaf and \textbf{nontrivial} otherwise. Removing an edge \(e\) of \(T\) induces the bipartition \(S_e\); the bipartitions induced this way are the \textbf{splits} of \(T\), of which \(n\) are trivial (pendant edges) and \(n-3\) nontrivial (internal edges). Two bipartitions are \textbf{compatible} if at least one of \(S\cap S'\), \(S\cap S'^c\), \(S^c\cap S'\), \(S^c\cap S'^c\) is empty, and \textbf{cross} otherwise. A set of nontrivial bipartitions is the split set of a tree if and only if it is pairwise compatible {[}19{]}, {[}20{]}; we call pairwise compatibility \textbf{laminarity}. There are \((2n-5)!!=1\cdot3\cdot5\cdots(2n-5)\) binary topologies on \(n\) labelled leaves: about
\(2\times10^6\) for \(n=10\), \(2\times10^{20}\) for \(n=20\) and \(9\times10^{36}\) for \(n=30\).

A \textbf{nearest-neighbour interchange (NNI)} across an internal edge \(uv\), with subtrees \(A,B\) hanging from \(u\) and \(C,D\) from \(v\), exchanges \(B\) with \(C\) or with \(D\). Each internal edge admits two NNIs, so a binary tree has \(2(n-3)\) NNI neighbours.

A symmetric nonnegative \(d\) is a \textbf{tree metric} if and only if it satisfies the \textbf{four-point condition} {[}19{]}: on every quartet \(\{i,j,k,l\}\), the larger two of \(d_{ij}+d_{kl}\), \(d_{ik}+d_{jl}\), \(d_{il}+d_{jk}\) are equal. A tree \textbf{resolves} the quartet as \(ab|cd\) when the path between \(a\) and \(b\) is disjoint from the path between \(c\) and \(d\). For the path lengths \(\tau\) of a binary tree, the two larger sums are then \(\tau_{ac}+\tau_{bd}=\tau_{ad}+\tau_{bc}\), and they exceed \(\tau_{ab}+\tau_{cd}\) by \(2m\ge2\), where \(m\) is the number of internal edges between the paths. We call this the \textbf{strong} four-point condition.

\subsection{2.2 The Balanced Minimum Evolution Problem}\label{the-balanced-minimum-evolution-problem}

A \textbf{dissimilarity matrix} (or distance matrix) on \([n]\) is a symmetric matrix \(D\) with zero diagonal and nonnegative entries, where \(D_{ij}\) estimates the evolutionary distance between taxa \(i\) and \(j\). Given \(D\), the minimum-evolution principle selects the tree of least estimated total length {[}21{]}. In the balanced version the length is \(L(T)=\sum_{i<j}D_{ij}w_{ij}\) with \(w_{ij}=2^{-\tau_{ij}}\) {[}1{]}, and the \textbf{Balanced Minimum Evolution Problem (BMEP)} is to minimize \(L\) over binary topologies.

\subsection{2.3 The weight space}\label{the-weight-space}

For a binary tree \(T\) write \(\tau_{ij}\in\{2,\dots,n-1\}\), \(w_{ij}=2^{-\tau_{ij}}\), and \(w_T=(w_{ij})_{i<j}\in\mathbb R^{\binom n2}\) for its \textbf{weight vector}. For any \(w\in\mathbb R^{\binom n2}\) and \(S\subseteq[n]\), the \textbf{cut value} of \(S\) is \(W[S]=\sum_{i\in S,\,j\notin S}w_{ij}\), and \(W_T[S]\) denotes the cut value for \(w=w_T\). Each edge \(e\) of \(T\) defines the bipartition \(S_e\) of the leaves on its two sides.

\begin{quote}
\textbf{Lemma 2.1 (edge cut values).} For every edge \(e\) of a binary tree \(T\), \(W_T[S_e]=\tfrac12\).
\end{quote}

\emph{Proof.} Let \(e=uv\), with leaf set \(A\) on the side of \(u\) and \(B\) on the side of \(v\). Rooted at \(u\), the side of \(A\) is a full binary tree (a single leaf if \(e\) is a pendant edge); let \(a_i\) be the number of edges from \(u\) to leaf \(i\in A\). Then \(\sum_{i\in A}2^{-a_i}=1\), by induction on \(|A|\): replacing a cherry at depth \(d+1\) by one leaf at depth \(d\) keeps the sum, since \(2\cdot2^{-(d+1)}=2^{-d}\). Likewise \(\sum_{j\in B}2^{-b_j}=1\), with \(b_j\) the number of edges from \(v\) to \(j\). For \(i\in A\) and \(j\in B\), \(\tau_{ij}=a_i+1+b_j\), so \[W_T[S_e]=\sum_{i\in A}\sum_{j\in B}2^{-a_i-1-b_j}
=\tfrac12\Bigl(\sum_{i\in A}2^{-a_i}\Bigr)\Bigl(\sum_{j\in B}2^{-b_j}\Bigr)=\tfrac12 .\qquad\square\]

For a pendant edge the lemma is the \textbf{Kraft equality} {[}6{]}, {[}7{]}: \(\sum_{j\ne i}w_{ij}=\tfrac12\) for every leaf \(i\), equivalently \(W[\{i\}]=\tfrac12\). Two further facts are used throughout.

\begin{itemize}
\tightlist
\item
  \textbf{Cut floor} {[}10{]}, {[}11{]}, {[}12{]}: \(W_T[S]\ge\tfrac12\), with equality if and only if \(S\) is a split of \(T\). Equality for splits is Lemma 2.1; strict inequality for the other bipartitions follows from Theorem 2 below (Appendix B).
\item
  \textbf{Edge decomposition.} \(\tau_{ij}\) is the number of edges of \(T\) separating \(i\) and \(j\), that is, \(\tau_{ij}=\sum_{e\in E(T)}[e\text{ separates }i\text{ and }j]\), where the Iverson bracket \([P]\) is \(1\) if \(P\) holds and \(0\) otherwise. Exchanging the order of summation, for every \(w\in\mathbb R^{\binom n2}\), \[\sum_{i<j}\tau_{ij}\,w_{ij}=\sum_{e\in E(T)}\sum_{i<j}[e\text{ separates }i\text{ and
  }j]\,w_{ij}=\sum_{e\in E(T)}W[S_e].\tag{2.1}\] For \(w=w_T\), Lemma 2.1 applied to the \(2n-3\) edges gives \(\sum_{i<j}\tau_{ij}(w_T)_{ij}=(2n-3)/2=n-\tfrac32\).
\end{itemize}

Since \(\log_2w_{ij}=-\tau_{ij}\), the last identity says that \(w_T\) lies on the \textbf{phylogenetic manifold} {[}16{]} \[g(w):=\sum_{i<j}w_{ij}\log_2w_{ij}=\tfrac32-n ,\] with the convention \(0\log0=0\). The function \(g\) is strictly convex on \(\mathbb R_{\ge0}^{\binom n2}\), since each term \(x\log_2x\) has second derivative \(1/(x\ln2)>0\) for \(x>0\). (The identity is usually written over ordered pairs, as \(\sum_i\sum_{j\ne i}\tau_{ij}2^{-\tau_{ij}}=2n-3\).) By Kraft, \(p^{(i)}=(2w_{ij})_{j\ne i}\) is a probability vector for each leaf \(i\), and its Shannon entropy \(H_i=-\sum_{j\ne i}p^{(i)}_j\log_2p^{(i)}_j\) {[}22, Ch. 2{]} satisfies \[\sum_{i=1}^n H_i=-n-4\,g(w).\tag{2.2}\] Minimizing \(g\) is therefore maximizing the \textbf{total leaf entropy} \(\sum_iH_i\), and the manifold can be equivalently written as \(\sum_iH_i=3(n-2)\). ``Entropy'' below always means \(\sum_iH_i\); the weights themselves sum to \(n/4\), not \(1\), and do not form a single probability
distribution. The information-theoretic reading of the BME objective and of the manifold is due to {[}23{]}.

The \textbf{BME polytope} \(P_{\mathrm{BME}}\) is the convex hull of the weight vectors \(w_T\) of all binary trees \(T\) on \([n]\). Several authors use the integer coordinate \(x_{ij}=2^{n-1}w_{ij}\in\{1,\dots,2^{n-3}\}\), in which the facets of Section 4 are usually stated.

\subsection{2.4 The two formulations}\label{the-two-formulations}

The \textbf{compact model} is the convex relaxation \[
\begin{aligned}
\min\ & \textstyle\sum_{i<j}D_{ij}w_{ij}\\
\text{s.t. }& \textstyle\sum_{j\ne i}w_{ij}=\tfrac12 && \forall i\in[n] &&\text{(Kraft)}\\
& 2^{-(n-1)}\le w_{ij}\le\tfrac14 && \forall i<j &&\text{(box; F1)}\\
& W[S]\ge\tfrac12 && \forall\,\emptyset\ne S\subsetneq[n] &&\text{(cut)}\\
& w_{ij}+w_{jk}-w_{ik}\le\tfrac14 && \forall\text{ distinct }i,j,k &&\text{(double cherry; F2)}\\
& g(w)\le\tfrac32-n && &&\text{(manifold)}\\
& \textstyle\sum_{\{i,j\}\in M}w_{ij}\ge C_n\,2^{-(n-1)} && \forall\text{ perfect matchings }M &&\text{(PM)}
\end{aligned}
\] The cut family can be separated exactly in polynomial time: a minimum cut of the complete graph on \([n]\) with edge weights \(w\) gives the most violated cut inequality. The manifold constraint is the convex half of \(g(w)=\tfrac32-n\), so every \(w_T\) satisfies it. PM is the perfect-matching family of {[}12{]}, prop. 11, with the constant \(C_n\) given there.

The compact model has no integer variables: a point is a tree because its \textbf{half-cuts} (the bipartitions with cut value \(\tfrac12\)) form a compatible split system, not because any variable is integral. Solving BMEP with it therefore needs a branch and bound of its own, which Section 7.2 describes.

The \textbf{distance-indexed model} {[}12{]}, {[}13{]} carries, for each pair \(i<j\) and each level \(\ell\in\{2,\dots,n-1\}\), a binary \textbf{level indicator} \(x^{(\ell)}_{ij}=[\tau_{ij}=\ell]\), so \(O(n^3)\) variables in all. The path lengths \(\tau_{ij}=\sum_\ell\ell\,x^{(\ell)}_{ij}\) and the weights \(w_{ij}=\sum_\ell2^{-\ell}x^{(\ell)}_{ij}\) are linear expressions in them. In the configuration we run (Section 7.1) the model is \[
\begin{aligned}
\min\ & \textstyle\sum_{i<j}D_{ij}\sum_{\ell}2^{-\ell}x^{(\ell)}_{ij}\\
\text{s.t. }& \textstyle\sum_{\ell}x^{(\ell)}_{ij}=1 && \forall i<j &&\text{(one length)}\\
& \textstyle\sum_{j\ne i}\sum_\ell2^{-\ell}x^{(\ell)}_{ij}=\tfrac12 && \forall i\in[n] &&\text{(Kraft)}\\
& \textstyle\sum_{i<j}\sum_\ell\ell\,2^{-\ell}x^{(\ell)}_{ij}=n-\tfrac32 && &&\text{(manifold)}\\
& \tau_{ik}+\tau_{kj}-\tau_{ij}\ge2 && \forall\text{ distinct }i,j,k &&\text{(strong triangle)}\\
& \tau_{ik}+\tau_{kj}-\tau_{ij}\in2\mathbb Z && \forall\text{ triples containing leaf }1 &&\text{(parity)}\\
& \text{quartet inequalities} && \forall\text{ quartets containing leaf }1 &&\text{(four-point)}\\
& x^{(\ell)}_{ij}\in\{0,1\}.
\end{aligned}
\] The parity rows use one integer variable per triple. The quartet inequalities use, for each quartet, three binary indicators of its resolutions, one of which is \(1\), and linear inequalities that impose the strong four-point condition for the selected resolution. Restricting them to the quartets that contain leaf \(1\) loses nothing: Catanzaro \emph{et al.} show that the strong four-point condition on these quartets implies it on every quartet, and, together with the strong triangle inequalities on the triples that contain leaf \(1\), implies the strong triangle inequalities on every triple {[}13{]}, thm. 2 and prop. 8. Appendix G gives a short proof of both. Catanzaro \emph{et al.} also define further families of valid inequalities for this model {[}13{]}: the \textbf{generalized Kraft} inequalities, which are the cut inequalities \(W[S]\ge\tfrac12\) written in the level indicators, and three further families. The \textbf{weak four-point} inequalities state that
\(2(\tau_{ui}+\tau_{uj}+\tau_{uk})-(\tau_{ij}+\tau_{ik}+\tau_{jk})\ge8\) for distinct leaves \(u,i,j,k\): the left side is twice the sum of the distances from \(u\) to the three paths between \(i\), \(j\) and \(k\), and in a binary tree these distances sum to at least \(4\), since \(u\) joins the three paths away from their common vertex, which already has degree \(3\) {[}13{]}, prop. 10. The \textbf{2-K-split} inequalities state that \(\sum_{k\in X}(\tau_{ik}+\tau_{jk})\ge\alpha_K\) for distinct leaves \(i,j\) and every set \(X\) of \(K\) other leaves, where \(\alpha_K\) is the largest constant valid for every tree, which can be computed in polynomial time {[}13{]}, prop. 11. The \textbf{circular-order} inequalities state that \(\tau_{i_1i_2}+\tau_{i_2i_3}+\dots+\tau_{i_{n-2}i_{n-1}}+
\tau_{i_{n-1}i_1}\ge4n-8\) for every leaf \(i_n\) and every cyclic order \((i_1,\dots,i_{n-1})\) of the other leaves: removing \(i_n\) and its pendant edge leaves a tree with \(2n-4\) edges, and a closed walk through its leaves traverses every edge at least twice {[}13{]}, prop. 9. The \textbf{four-point cuts} are the quartet inequalities above, added when violated. The two models solve the same problem and differ in which constraints are linear; Section 6 makes this precise. Before that, the next section shows that in the weight space a tree is determined by its splits.

\section{3. Splits, cut values and the weight vector}\label{splits-cut-values-and-the-weight-vector}

This section contains the result the rest of the paper builds on: the splits of a tree, written as linear equalities in \(w\), together with the convex half of the phylogenetic manifold, leave exactly one point, the tree's weight vector (Theorem 1). Section 3.2 shows that the equalities cannot be weakened to inequalities, and Section 3.3 describes the cut values of the bipartitions that are not splits.

\subsection{3.1 Pinning a fixed tree}\label{pinning-a-fixed-tree}

For a binary topology \(T\) let \[\mathcal A_T=\Bigl\{w\in\mathbb R_{\ge0}^{\binom n2}:\ \textstyle\sum_{j\ne i}w_{ij}=\tfrac12\ \forall i,\quad
W[S]=\tfrac12\ \forall\text{ nontrivial splits }S\text{ of }T\Bigr\}.\] This is a polytope, since Kraft and \(w\ge0\) give \(w_{ij}\le\tfrac12\). It is defined by \(2n-3\) equalities, so for \(n\ge5\) it is far from a single point.

\begin{quote}
\textbf{Theorem 1 (pinning).} For every \(w\in\mathcal A_T\), \[g(w)\;=\;\bigl(\tfrac32-n\bigr)+\sum_{i<j}w_{ij}\log_2\frac{w_{ij}}{(w_T)_{ij}},\] and the sum is \(\ge0\) with equality if and only if \(w=w_T\). Consequently \(w_T\) is the unique minimizer of \(g\) over \(\mathcal A_T\) (the unique maximizer of \(\sum_iH_i\), with value \(3(n-2)\)), and \(\mathcal A_T\cap\{g\le\tfrac32-n\}=\{w_T\}\).
\end{quote}

\emph{Proof.} Let \(w\in\mathcal A_T\). By (2.1), \(\sum_{i<j}\tau_{ij}w_{ij}=\sum_eW[S_e]\); each pendant edge contributes a Kraft row and each internal edge a split equality, each equal to \(\tfrac12\), so the sum is \(n-\tfrac32\). Since \(\log_2(w_T)_{ij}=-\tau_{ij}\), \[g(w)-\bigl(\tfrac32-n\bigr)=\sum w_{ij}\log_2w_{ij}+\sum\tau_{ij}w_{ij}
=\sum w_{ij}\log_2\frac{w_{ij}}{(w_T)_{ij}} .\] Summing the Kraft rows gives \(\sum_{i<j}w_{ij}=n/4=\sum_{i<j}(w_T)_{ij}\). The log-sum inequality {[}22{]}, thm. 2.7.1 then gives \(\sum w\log_2(w/w_T)\ge(\sum w)\log_2(\sum w/\sum w_T)=0\), with equality if and only if \(w\) is proportional to, hence equal to, \(w_T\). The convention \(0\log0=0\) is consistent because \(w_T>0\). \(\blacksquare\)

The content of Theorem 1 is that \(2n-3\) linear equalities and one convex inequality, in \(\binom n2\) variables and without integrality, describe a fixed tree exactly. It does not give an algorithm, because \(\mathcal A_T\) presupposes \(T\); the hardness of the problem therefore lies precisely in determining the splits, which determine the tree.

\textbf{Remark (flatness).} The function \(g\) is strictly convex, with Hessian \(\operatorname{diag}\bigl(1/(w_{ij}\ln2)\bigr)\), and on \(\mathcal A_T\) it attains its minimum at \(w_T\), so \(g-g(w_T)\) grows quadratically near \(w_T\). Kelley's cutting-plane method {[}24{]} imposes a convex constraint through finitely many of its tangent hyperplanes, added where the current point violates it. Relaxing the manifold to \(g\le\tfrac32-n+\varepsilon\), as any finite set of such tangents does, admits points at distance \(\Theta(\sqrt\varepsilon)\) from \(w_T\), so a solver that pins the tree through tangent cuts carries an \(O(\sqrt\varepsilon)\) objective error.

\subsection{3.2 Why equalities and not inequalities: the star point}\label{why-equalities-and-not-inequalities-the-star-point}

The pinning is lost if the split equalities are replaced by the cut inequalities \(W[S]\ge\tfrac12\). The \textbf{star point} \(w^\star_{ij}=\frac1{2(n-1)}\), the balanced weights of the star tree, satisfies Kraft and every cut inequality, since \(W^\star[S]=\frac{|S|(n-|S|)}{2(n-1)}\ge\tfrac12\). For every integer \(n\ge4\) it also satisfies \[g(w^\star)=-\tfrac n4\bigl(1+\log_2(n-1)\bigr)\;<\;\tfrac32-n\] (Appendix A). So the star lies strictly inside \(\{\text{Kraft},\ W[S]\ge\tfrac12\ \forall S,\ g\le\tfrac32-n\}\). That relaxation does not pin a tree, and the entropy constraint alone cannot separate trees from non-trees. The star is excluded from \(\mathcal A_T\) only because its nontrivial cuts exceed \(\tfrac12\).

\subsection{3.3 The cut-value ladder and an invariant count}\label{the-cut-value-ladder-and-an-invariant-count}

Let us fix a binary tree \(T\) and a bipartition \(S\), and colour each leaf \(S\) or \(S^c\). An edge \(e\) is a \textbf{boundary edge} if both components of \(T-e\) contain leaves of both colours; \(B(S)\) is the set of boundary edges. Lemma B.1 shows that \(B(S)\) is either empty or a subtree of \(T\) spanning internal vertices only. For each vertex \(u\) of \(B(S)\), the components of \(T-u\) that contain no boundary edge are the \textbf{pendant subtrees} at \(u\), and each is monochromatic. A \textbf{path boundary} of length \(k\) is a path \(v_0v_1\cdots v_k\). Each of its end vertices carries two pendant subtrees, necessarily of opposite colours, and each interior vertex carries exactly one. The colouring is \textbf{constant} if all the interior pendant subtrees have the same colour.

\begin{quote}
\textbf{Theorem 2 (cut-value ladder).} For every binary tree \(T\) and bipartition \(S\):

\begin{enumerate}
\def\labelenumi{\arabic{enumi}.}
\tightlist
\item
  \(B(S)=\emptyset\) if and only if \(S\) is a split of \(T\), and then \(W_T[S]=\tfrac12\).
\item
  \(|B(S)|=1\) if and only if \(S\) is a split of a tree one NNI away from \(T\) but not a split of \(T\), and then \(W_T[S]=\tfrac34\). There are exactly \(2(n-3)\) such bipartitions, two per internal edge, in bijection with the NNI moves of \(T\).
\item
  If \(B(S)\) is a path of length \(k\ge2\), then \(W_T[S]=1-2^{-(k+1)}+\tfrac14P\), where \(P=\sum_{1\le i<j\le k-1,\ c_i\ne c_j}2^{-(j-i)}\ge0\) sums over pairs of interior pendant subtrees of opposite colours. In particular \(W_T[S]=1-2^{-(k+1)}\) (the rungs \(\tfrac78,\tfrac{15}{16},\dots\)) if and only if the colouring is constant, and \(W_T[S]>1\) otherwise.
\item
  If \(B(S)\) is not a path, then \(W_T[S]\ge\tfrac98\), and equality is attained.
\end{enumerate}

Hence \(W_T[S]\in\{\tfrac12\}\cup\{\tfrac34\}\cup[\tfrac78,\infty)\), and the open intervals \((\tfrac12,\tfrac34)\) and \((\tfrac34,\tfrac78)\) contain no cut value of any tree.
\end{quote}

The proof (Appendix B) rests on one identity. Seen from its attachment vertex, each pendant subtree carries total Pauplin weight \(\tfrac12\). Hence \(4W_T[S]\) is the sum of \(2^{-d}\) over pairs of opposite-coloured pendant subtrees, where \(d\) is their distance along \(B(S)\) (Lemma B.3).

\begin{quote}
\textbf{Theorem 3 (invariant count).} For every binary tree \(T\) on \(n\ge4\) leaves, \[\bigl|\{\text{bipartitions }S:\ W_T[S]\le1\}\bigr|=2n^2-10n+15,\] counting \(\{S,S^c\}\) once and including the \(n\) trivial splits.
\end{quote}

\emph{Proof sketch} (full proof in Appendix B). By Theorem 2, \(W_T[S]\le1\) if and only if \(B(S)\) is empty or a path with constant colouring. The empty case gives the \(2n-3\) splits. A path is determined by its two end vertices, an unordered pair of distinct internal vertices, of which there are \(\binom{n-2}2\). Conversely, every such pair with every admissible colouring yields a bipartition whose boundary is exactly that path. A path of length \(1\) admits \(2\) bipartitions (the NNI pair). A path of length \(k\ge2\) admits \(4\): \(2\times2\) end colourings, \(2\) choices of constant interior colour, halved for complementation. The \(n-3\) internal edges are the paths of length \(1\), so the count is \((2n-3)+2(n-3)+4\bigl[\binom{n-2}2-(n-3)\bigr]=2n^2-10n+15\). \(\blacksquare\)

\textbf{Corollary.} By Kraft, every pair cut satisfies \(W[\{i,j\}]=W[\{i\}]+W[\{j\}]-2w_{ij}=1-2w_{ij}<1\). Removing the \(\binom n2\) pair cuts and the \(n\) singletons from the count of Theorem 3 leaves exactly \(\tfrac32(n-2)(n-5)\) bipartitions with both blocks of size \(\ge3\) and \(W_T[S]\le1\), again independent of \(T\) (\(n\ge5\)). The subtraction needs the \(\binom n2\) pair cuts to be distinct bipartitions, which holds for \(n\ge5\); at \(n=4\) each \(2|2\) bipartition is two pair cuts, and the count of blocks of size \(\ge3\) is zero (as it is at \(n=5\)).

The individual rungs above \(\tfrac34\) vary with the topology; only the total up to \(1\) is invariant.

\textbf{Worked example} (\(n=6\)). We take the \textbf{caterpillar} (the tree whose internal vertices form a path) with internal path \(u_1u_2u_3u_4\), cherries \(\{1,2\}\) at \(u_1\) and \(\{5,6\}\) at \(u_4\), and leaves \(3\), \(4\) at \(u_2\), \(u_3\) (Figure 1a). Table 2 lists four bipartitions.

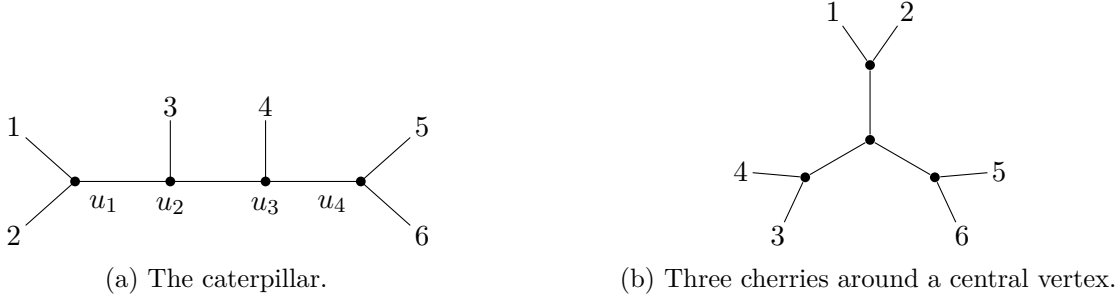
\begin{figure}[ht]
\centering
\tikzset{iv/.style={circle,fill,inner sep=1.3pt},lf/.style={inner sep=1.5pt}}
\begin{subfigure}[b]{0.48\textwidth}\centering
\begin{tikzpicture}[scale=0.9]
\foreach \x/\l/\p in {0/u_1/below right,1.4/u_2/below,2.8/u_3/below,4.2/u_4/below left}{\node[iv,label=\p:{$\l$}] at (\x,0) {};}
\draw (0,0)--(4.2,0);
\node[lf] (a) at (-0.9,0.8) {$1$}; \node[lf] (b) at (-0.9,-0.8) {$2$};
\node[lf] (c) at (1.4,1.1) {$3$}; \node[lf] (d) at (2.8,1.1) {$4$};
\node[lf] (e) at (5.1,0.8) {$5$}; \node[lf] (f) at (5.1,-0.8) {$6$};
\draw (0,0)--(a) (0,0)--(b) (1.4,0)--(c) (2.8,0)--(d) (4.2,0)--(e) (4.2,0)--(f);
\end{tikzpicture}
\caption{The caterpillar.}
\end{subfigure}\hfill
\begin{subfigure}[b]{0.48\textwidth}\centering
\begin{tikzpicture}[scale=0.9]
\node[iv] (o) at (0,0) {};
\foreach \a/\i/\j in {90/1/2,210/3/4,330/5/6}{
  \node[iv] (v\i) at (\a:1.1) {};
  \draw (o)--(v\i);
  \node[lf] (l\i) at ($(\a:1.1)+(\a+35:0.95)$) {$\i$};
  \node[lf] (l\j) at ($(\a:1.1)+(\a-35:0.95)$) {$\j$};
  \draw (v\i)--(l\i) (v\i)--(l\j);}
\end{tikzpicture}
\caption{Three cherries around a central vertex.}
\end{subfigure}
\caption{The two 6-leaf trees of the worked example in Section 3.3.}
\label{fig:six-leaf}
\end{figure}

\textbf{Table 2. Boundaries and cut values of four bipartitions of the 6-leaf caterpillar.}

{\def\LTcaptype{none} 
\begin{longtable}[]{@{}llll@{}}
\toprule\noalign{}
\(S\) & \(B(S)\) & case & \(W_T[S]\) \\
\midrule\noalign{}
\endhead
\bottomrule\noalign{}
\endlastfoot
\(\{1,2\}\) & \(\emptyset\) & split & \(1/2\) \\
\(\{1,3\}\) & \(u_1u_2\) & one NNI & \(3/4\) \\
\(\{1,3,4,5\}\) & \(u_1\cdots u_4\) & constant interior & \(15/16\) \\
\(\{1,3,5\}\) & \(u_1\cdots u_4\) & \(3\in S\), \(4\notin S\), so \(P=1/2\) & \(17/16\) \\
\end{longtable}
}

For a branching boundary, we take the tree with cherries \(\{1,2\},\{3,4\},\{5,6\}\) around a central vertex (Figure 1b). Then \(S=\{1,3,5\}\) has all three internal edges in \(B(S)\), and \(W_T[S]=3\cdot\tfrac14+6\cdot\tfrac1{16}=\tfrac98\), the minimum in part 4.

The floor \(W\ge\tfrac12\) and the dichotomy between \(=\tfrac12\) and \(>\tfrac12\) are published (Section 2.3). Haws, Hodge and Yoshida {[}11{]}, cor. 1 show that NNI neighbours are adjacent vertices of \(P_{\mathrm{BME}}\), but the correspondence between the NNIs and the cuts of value \(\tfrac34\) appears to be new, as do the rest of the ladder and the invariant count.

Theorems 1 and 2 describe a tree once its splits are known; neither says how to find the right splits. A solver searches for them over an outer relaxation of the BME polytope, so the next section reviews the inequalities known to be valid for it.

\section{\texorpdfstring{4. Facets of the BME polytope for small \(n\)}{4. Facets of the BME polytope for small n}}\label{facets-of-the-bme-polytope-for-small-n}

In this section we review the known valid inequalities of the BME polytope and compute its facets for \(n=6\), to see how much of the polytope they describe. An \textbf{outer relaxation} of \(P_{\mathrm{BME}}\) is a set that contains it, described by inequalities valid for every \(w_T\); the feasible set of the compact model is one. The objective is linear in \(w\), so the gap between the optimum over an outer relaxation and the optimum over \(P_{\mathrm{BME}}\) comes entirely from the points of the relaxation outside \(P_{\mathrm{BME}}\). Table 3 lists the known families of valid inequalities, translated into \(w\)-space and numbered F1--F13 as in {[}14{]}, table 1.

\textbf{Table 3. Known inequalities of the BME polytope in \(w\)-space.} For each inequality we record whether the source proves it is facet-defining (facet) or we only know it is valid (valid); whether our experiments ever separate it (used); and whether it is in the default cut set (default).

{\def\LTcaptype{none} 
\begin{longtable}[]{@{}lllll@{}}
\toprule\noalign{}
\# & source & \(w\)-form & facet & used \\
\midrule\noalign{}
\endhead
\bottomrule\noalign{}
\endlastfoot
F1 & {[}9{]} & \(w_{ij}\ge2^{-(n-1)}\) & facet & default \\
F2 & {[}9{]} & \(w_{ij}+w_{jk}-w_{ik}\le\tfrac14\) & valid & default \\
F3 & {[}9{]} & \(w_{ij}+w_{ik}+w_{jk}\le\tfrac12\) & valid & implied \\
F4 & {[}12{]} & \(2^{n-5}(w_{i_1i_2}+w_{i_1i_3})+w_{i_2i_3}\ge\tfrac5{16}\) & valid & optional \\
F5 & {[}12{]} & \(2^{n-4}(w_{i_1i_2}+w_{i_1i_3})+w_{i_2i_3}\ge\tfrac12\) & valid & optional \\
F6 & {[}12{]} & \(w_{i_3i_4}\le2^{n-4}w_{i_1i_2}\) & facet & optional \\
F7--F12 & {[}12{]} & 5-, 6- and \(n\)-leaf families & some & F7 optional; PM default \\
F13 & {[}12{]} & \(w_{i_1i_2}+w_{i_1i_3}+w_{i_2i_3}\ge(3+\rho)2^{k-1}/2^{n-1}\) & valid & no \\
\end{longtable}
}

In F6 the four leaves are distinct; F7--F12 include the perfect-matching, star and cycle inequalities. F3 is implied by Kraft and the cut inequalities: by Kraft, \(W[\{i,j,k\}]=\tfrac32-2(w_{ij}+w_{ik}+w_{jk})\), so the cut inequality \(W[\{i,j,k\}]\ge\tfrac12\) is F3. Being implied in this way does not prevent it from being a facet: at \(n=6\) it is the balanced \((3,3)\) cut inequality, which does define facets of \(P_6\) (below).

The named families are a small fraction of the facets. We converted the 105 vertices of \(P_6\) to a facet description with cddlib {[}25{]}, after using the 6 Kraft equalities to eliminate the coordinates \(w_{ij}\) of the six pairs at circular distance 2 on the cycle \(1,2,\dots,6\), which leaves 9 coordinates, and obtained 90,262 facets. Lifting F1, F2 and F6 into these coordinates matches 165 of them (F1: 15, F2: 60, F6: 90), or \(0.18\%\); in these coordinates \(90\%\) of the facets involve all 9. Among the cut inequalities only the balanced \((3,3)\) family defines facets of \(P_6\): \((1,5)\)-cuts are Kraft equalities and \((2,4)\)-cuts reduce to \(w_{ij}\le\tfrac14\). We measure what the named families add to the bound in Section 8.2. Since the known facets describe the polytope only in part, the next section approaches it from the other side, through extended formulations whose variables encode the splits themselves.

\section{5. A polynomial laminarity, and splits via membership}\label{a-polynomial-laminarity-and-splits-via-membership}

By Theorem 1, finding a BME tree amounts to finding its splits. We can think of this search as determining an implicit vector \(x\) indexed by the nontrivial bipartitions, where \(x_S=1\) if and only if the tree we are looking for has \(S\) as a split. Additionally, we can introduce auxiliary variables \(s_{ik}\) for each leaf \(i\) and each of the \(n-3\) internal splits of the tree, numbered \(k=1,\dots,n-3\), with \(s_{ik}=1\) if and only if leaf \(i\) lies on a chosen side of split \(k\). Both lead to extended formulations of BMEP: the variables \(x_S\), whose compatibility constraints we make polynomial in number, in Section 5.1, and the variables \(s_{ik}\) in Section 5.2. Although neither extended formulation turned out to be practical (Section 8.4), we nevertheless prove some of their theoretical properties.

\subsection{5.1 Incompatibility and a polynomial clique cover}\label{incompatibility-and-a-polynomial-clique-cover}

Let \(G\) be the \textbf{crossing graph} on nontrivial bipartitions, with an edge between every crossing pair. A clique of \(G\) is a pairwise-crossing family, and a tree contains at most one split from it. For split indicators \(x_S\) the inequality \(\sum_{S\in K}x_S\le1\) is therefore valid, and it implies every pairwise inequality \(x_S+x_{S'}\le1\) inside \(K\). An \textbf{edge clique cover} of \(G\), a family of cliques such that every edge lies in at least one of them, therefore replaces the \(\Theta(4^n)\) pairwise laminarity constraints.

\textbf{Fixed-leaf representation.} For a leaf \(x\) write \(A_x(S)\) for the block of \(S\) containing \(x\). Since \(A_x(S)\cap A_x(S')\) always contains \(x\), the bipartitions \(S\) and \(S'\) are compatible if and only if \(A_x(S)\subseteq A_x(S')\), \(A_x(S')\subseteq A_x(S)\), or \(A_x(S)\cup A_x(S')=[n]\).

\begin{quote}
\textbf{Proposition 4 (polynomial laminarity).} (a) For distinct leaves \(x,y\) and sizes \(2\le j\le k\) with \(j+k\le n\), let \[Q^{j,k}_{x,y}=\{S:|A_x(S)|=j,\ y\in A_x(S)\}\cup\{S:|A_x(S)|=k,\ y\notin A_x(S)\}.\] Each \(Q^{j,k}_{x,y}\) is a clique of \(G\), and these \(O(n^4)\) cliques cover every edge of \(G\). (b) Every edge clique cover of \(G\) has at least \(8\,(1-e^{-n/16}-2^{-n/4})\binom m2\) cliques, where \(m\) is the number of integers \(i\) with \(\max(2,n/4)\le i<n/2\); in particular, at least \((1-o(1))\,n^2/4\).
\end{quote}

\emph{Proof of (a).} \emph{Clique.} Two members from the first half have distinct \(x\)-blocks of size \(j\), so neither contains the other, and both blocks contain \(x\) and \(y\), so their union has at most \(2j-2<n\) elements (as \(j\le n/2\)). Two members from the second half have distinct \(x\)-blocks of size \(k\), and their union misses \(y\). For blocks \(A\ni y\) from the first half and \(A'\not\ni y\) from the second, \(y\in A\setminus A'\); \(A'\) is not contained in \(A\), since it would then lie in \(A\setminus\{y\}\), which has \(j-1<k\) elements; and their union has at most \(j+k-1\le n-1\) elements, because \(x\in A\cap A'\). By the fixed-leaf criterion, every pair crosses. \emph{Cover.} Let \(S,S'\) cross, with quadrants \(Q_1=S\cap S'\), \(Q_2=S\setminus S'\), \(Q_3=S'\setminus S\) and \(Q_4=[n]\setminus(S\cup S')\), all nonempty. Choosing the reference leaf in \(Q_1\) gives \(|A|+|A'|=n+|Q_1|-|Q_4|\); choosing it in \(Q_4\) gives \(n+|Q_4|-|Q_1|\). We take \(x\) in
the smaller of \(Q_1\) and \(Q_4\), so that \(|A|+|A'|\le n\), and name the two bipartitions so that \(|A|\le|A'|\). The \(x\)-blocks are not nested, so we can take \(y\in A\setminus A'\), and then \(S\) and \(S'\) both lie in \(Q^{|A|,|A'|}_{x,y}\). \emph{Count.} There are \(n(n-1)\) ordered pairs \((x,y)\) and \(O(n^2)\) size pairs. \(\blacksquare\)

The condition \(j+k\le n\) cannot be dropped: without it the families are not cliques in general. At \(n=6\), \(Q^{4,4}_{1,2}\) contains \(A=\{1,2,4,5\}\) and \(A'=\{1,3,4,6\}\), whose union is \([6]\), so the corresponding splits \(\{3,6\}\) and \(\{2,5\}\) are compatible.

Part (b) is a counting argument, proved in Appendix H. The smaller blocks of the bipartitions in a clique pairwise intersect and none contains another, so by an inequality of Bollobás {[}26{]} and of Greene, Katona and Kleitman {[}27{]}, a clique has at most \(\beta_i\binom{n-1}{i-1}\) smaller blocks of each size \(i\), with \(\sum_i\beta_i\le1\). For sizes \(n/4\le i<j<n/2\) almost every pair of an \(i\)-set and a \(j\)-set crosses, and a clique covers at most a share of about \(\beta_i\beta_j/4\) of these pairs. Each such size pair must be covered in full, while \(\sum_{i<j}\beta_i\beta_j\le\tfrac12\) for every clique, and there are about \(n^2/32\) size pairs.

\textbf{Sizes.} The cover of part (a) has \(O(n^4)\) cliques. In split-indicator form these are \(O(n^4)\) constraints over \(\Theta(2^n)\) variables \(x_S\), each of exponential support. Aggregating each half into an indicator, \(a^+_{x,y,t}\) (``\(T\) has a split whose \(x\)-block has size \(t\) and contains \(y\)'') or \(a^-_{x,y,t}\) (the same with \(y\) excluded), gives \(O(n^3)\) variables and \(O(n^4)\) two-term constraints \(a^+_{x,y,j}+a^-_{x,y,k}\le1\). An integer program for set cover over the maximal cliques gives the minimum edge clique cover \(1,5,21,49\) for \(n=4,\dots,7\). A subfamily of the cover of part (a) attains these values, and at \(n=8\) it has \(116\) cliques, while every edge clique cover has at least \(88\) (Appendix H), so it is within a factor of \(1.32\) of the minimum. Between the bounds \(\Omega(n^2)\) and \(O(n^4)\), the growth rate of the minimum is open.

\textbf{Integrality.} The clique relaxation is not integral. \(G\) contains induced 5-cycles, for example at \(n=6\) the bipartitions \(\{2,3,6\},\{5,6\},\{2,6\},\{2,3\},\{3,4\}\), where consecutive ones cross and non-consecutive ones are compatible, so \(G\) is not a perfect graph. LPs over the cover with random objectives reach fractional optima for every \(n=5,\dots,8\). Proposition 4 gives a compact description of compatibility for problems whose objective is linear in the split indicators, such as maximum-weight compatible splits. We revisit it as part of a BME model in Section 8.4.

\subsection{5.2 Side-membership variables}\label{side-membership-variables}

The \textbf{membership model} encodes each of the \(n-3\) internal splits by a binary side-membership vector \(s_{ik}\in\{0,1\}\), with \(s_{ik}=1\) if and only if leaf \(i\) is on the reference side of split \(k\). Split \(k\) separates leaves \(i\) and \(j\) if and only if exactly one of \(s_{ik}\) and \(s_{jk}\) is \(1\), that is, if and only if their exclusive or (XOR) \(B_{ijk}=s_{ik}\oplus s_{jk}\) is \(1\). The four inequalities \[B_{ijk}\le s_{ik}+s_{jk},\quad B_{ijk}\le2-s_{ik}-s_{jk},\quad B_{ijk}\ge s_{ik}-s_{jk},\quad B_{ijk}\ge s_{jk}-s_{ik}\] define the convex hull of the XOR relation and are its facets. Then \(\tau_{ij}-2=\sum_kB_{ijk}\). We linearize the product \(P_{ijk}=B_{ijk}w_{ij}\), with \(w_{ij}\in[L,U]=[2^{-(n-1)},\tfrac14]\), by the McCormick inequalities {[}28{]} \[LB_{ijk}\le P_{ijk}\le UB_{ijk},\qquad w_{ij}-U(1-B_{ijk})\le P_{ijk}\le w_{ij}-L(1-B_{ijk}),\] which are exact at binary \(B\). The cut coupling \(\sum_{i<j}P_{ijk}=\tfrac12\) is the split equality
of Theorem 1 for split \(k\). We enforce laminarity pairwise, with one binary variable per pair of splits that selects between nesting and disjointness. Symmetry breaking takes the minority side as the reference side \(A_k\) and, for even \(n\), places leaf \(1\) outside every split of size \(n/2\); it orders the splits by size, requires the first two to be disjoint cherries (every tree with \(n\ge6\) has two), and lists splits of equal size in strictly decreasing lexicographic order of their membership vectors: \[v_k-v_{k+1}\ge1-2^n\bigl(|A_{k+1}|-|A_k|\bigr),\qquad v_k=\sum_i2^{\,n-i}s_{ik},\] which forces \(v_k>v_{k+1}\) when \(|A_k|=|A_{k+1}|\) and is slack otherwise.

The leaf-\(1\) condition is needed for correctness as well as symmetry. Without it, two complementary balanced splits pass every compatibility test, which gives duplicate splits and non-trees (at \(n=8\): 17,535 split systems instead of 10,395). The lexicographic order removes the remaining multiplicity: without it a tree with \(c_m\) splits of size \(m\) is represented \(\prod_mc_m!\) times.

\begin{quote}
\textbf{Conjecture 1.} With these symmetry-breaking constraints, the integral points of the membership model are in one-to-one correspondence with the \((2n-5)!!\) binary topologies, for every \(n\).
\end{quote}

Complete enumeration confirms Conjecture 1 for \(n\le8\): the model has exactly 105, 945 and 10,395 integral points for \(n=6,7,8\), one for each tree.

\subsection{5.3 Cut-polytope inequalities on the membership layer}\label{cut-polytope-inequalities-on-the-membership-layer}

Each column \(B_{\cdot\cdot k}\) is a cut vector, so the cut-polytope triangle inequalities \(B_{ijk}\le B_{ipk}+B_{jpk}\) and \(B_{ijk}+B_{ipk}+B_{jpk}\le2\) are valid for each split. Summing the first over \(k\) and using \(\tau_{ij}-2=\sum_kB_{ijk}\) gives the strong triangle inequality \(\tau_{ip}+\tau_{jp}-\tau_{ij}\ge2\) of {[}12{]}. The per-split form therefore implies the aggregated one, and it is strictly stronger as a constraint on \(B\): with two splits, \((B_{ij1},B_{ip1},B_{jp1})=(1,0,0)\) and \((B_{ij2},B_{ip2},B_{jp2})=(0,1,1)\) satisfy the aggregated inequality (\(3+3-3\ge2\)) but violate the per-split inequality for split 1. The odd-cycle inequalities of the cut polytope can be separated in polynomial time {[}29{]}. We revisit both families in Section 8.4. The extended formulations of this section keep the objective linear in \(w\), but the condition that makes a matrix a tree metric, the four-point condition, is not linear in \(w\); the next section compares which
constraints are linear in \(w\) and which in \(\tau\).

\section{\texorpdfstring{6. Linearity in \(w\) versus linearity in \(\tau\)}{6. Linearity in w versus linearity in \textbackslash tau}}\label{linearity-in-w-versus-linearity-in-tau}

This section compares which constraints are linear in the two coordinate systems \(w_{ij}=2^{-\tau_{ij}}\) and \(\tau_{ij}=-\log_2w_{ij}\), with no auxiliary variables. It describes a trade-off. It is not a lower bound on extended formulations, which may add variables freely; we return to that question in Section 9.

\subsection{\texorpdfstring{6.1 The four-point condition in \(w\)}{6.1 The four-point condition in w}}\label{the-four-point-condition-in-w}

For the resolution \(ab|cd\) of a quartet, the strong four-point condition in \(\tau\) is the pair of linear constraints \[\tau_{ac}+\tau_{bd}=\tau_{ad}+\tau_{bc},\qquad \tau_{ac}+\tau_{bd}\ge\tau_{ab}+\tau_{cd}+2 .\] In \(w\) these become \[w_{ac}w_{bd}=w_{ad}w_{bc},\qquad w_{ab}w_{cd}\ge4\,w_{ac}w_{bd},\] so for a fixed resolution the feasible set in \(w\) is non-convex. In either coordinate system the condition is a disjunction over the three resolutions. The standard linear outer approximation of a product \(p=uv\) over a box \([L,U]^2\), the McCormick envelope {[}28{]}, is weakest at the centre of the box, where it underestimates \(p\) by \(\bigl(\tfrac{U-L}2\bigr)^2\). With \(L=2^{-(n-1)}\approx0\) and \(U=\tfrac14\) this is \(\tfrac1{64}\), the whole of \(p\) at \(u=v=\tfrac18\), so the envelope reduces to \(p\ge0\).

\subsection{6.2 The trade-off}\label{the-trade-off}

\textbf{Table 4. Linearity of the main objects in each coordinate system, with no auxiliary variables.}

{\def\LTcaptype{none} 
\begin{longtable}[]{@{}lll@{}}
\toprule\noalign{}
object & in \(w\) & in \(\tau\) \\
\midrule\noalign{}
\endhead
\bottomrule\noalign{}
\endlastfoot
BME objective \(\sum D_{ij}w_{ij}\) & linear & convex (\(D\ge0\)), nonlinear \\
Kraft (one leaf) & linear & nonlinear equality \\
manifold (an equality on trees) & only the half \(g\le\frac32-n\) is convex & nonlinear \\
strong triangle & \(w_{ij}\ge4\,w_{ik}w_{kj}\), non-convex & linear \\
four-point (fixed resolution) & bilinear, non-convex & linear \\
2-K-split, circular order & non-convex & linear \\
cut \(W[S]\ge\frac12\) (generalized Kraft) & linear & nonlinear \\
\end{longtable}
}

In \(w\) the objective and the cut and Kraft constraints are linear, but the quartet and triangle conditions are not; in \(\tau\) the reverse holds. The distance-indexed model gets both by lifting. Its level indicators \(x^{(\ell)}_{ij}\) make the objective (\(\sum_\ell2^{-\ell}x^{(\ell)}\)) and the \(\tau\)-conditions (\(\tau=\sum_\ell\ell
x^{(\ell)}\)) linear at once, at the cost of \(\Theta(n^3)\) variables. The same lift makes the whole manifold linear: by (2.1) every tree satisfies the equality \(\sum_{i<j}\sum_\ell\ell\,2^{-\ell}x^{(\ell)}_{ij}=n-\tfrac32\), whereas in \(w\) only its convex half \(g(w)\le\tfrac32-n\) can be imposed, as a convex constraint. The claim that no coordinate system makes both linear therefore holds only for the two unlifted systems. We revisit the lifted manifold in Section 8.7.

The generalized Kraft cut of the distance-indexed model builds \(w_{ij}=\sum_\ell2^{-\ell}x^{(\ell)}_{ij}\) and imposes \(W[S]\ge\tfrac12\), which is linear in \(w\) and is exactly the min-cut family of the compact model. Its 2-K-split, circular-order and four-point cuts are linear in \(\tau\). Rewritten in \(w\) they take the form \(\sum c_{ij}(-\log_2w_{ij})\ge b\) with \(c\ge0\). These are superlevel sets of convex functions and are not convex in \(w\), so none has an exact convex counterpart in \(w\)-coordinates. Valid convex outer approximations exist (for example tangent cuts of \(P_{\mathrm{BME}}\)), but they are relaxations.

\subsection{6.3 Sufficiency of the path-length conditions}\label{sufficiency-of-the-path-length-conditions}

Catanzaro, Pesenti and Wolsey {[}12{]} give necessary conditions for an integer matrix \(\tau\) to be the path-length matrix of a binary tree: (4) zero diagonal, (5) symmetry, (6) strong triangle \(\tau_{ik}+\tau_{kj}-\tau_{ij}\ge2\), (7) \(\tau_{ij}\in\{2,\dots,n-1\}\), (8) Kraft, (9) manifold. Adding the strong four-point condition (10) makes them sufficient: a matrix satisfying (4)--(10) is the path-length matrix of a binary tree {[}12{]}, {[}13{]}, thm. 1. Their open question is whether (4)--(9) already suffice, that is, whether (10) follows from the rest.

\begin{quote}
\textbf{Theorem 5 (computer-assisted).} (4)--(9) imply (10) for every \(n\le11\). For \(n=12\) they do not, and up to relabelling of the leaves there is exactly one \(12\times12\) matrix satisfying (4)--(9) but not (10).
\end{quote}

\textbf{The counterexample (cherry lift).} Let \[r=\begin{pmatrix}0&2&3&4&5&5\\2&0&4&3&5&5\\3&4&0&3&3&4\\4&3&3&0&4&3\\5&5&3&4&0&2\\5&5&4&3&2&0\end{pmatrix}.\] Each row of \(r\) satisfies \(\sum_{l\ne k}2^{-r_{kl}}=\tfrac12\), and \(r\) satisfies the weak triangle inequality \(r_{kl}+r_{lm}\ge r_{km}\) but is not a tree metric. We replace each of the six nodes by a cherry: leaves \(2a-1,2a\) (\(a=1,\dots,6\)) form cherry \(a\), and \[\tau_{pq}=\begin{cases}2,&p,q\text{ in the same cherry},\\ r_{ab}+2,&p\in\text{cherry }a,\ q\in\text{cherry }b,\ a\ne b.\end{cases}\] Kraft lifts row by row: \(\tfrac14+2\cdot\tfrac14\sum_b2^{-r_{ab}}=\tfrac12\). The shift by \(2\) turns the weak triangle into the strong one, since \((r_{ac}+2)+(r_{cb}+2)-(r_{ab}+2)\ge2\). The manifold holds by direct computation. For the four-point condition, we take the quartet \(\{1,2,3,5\}\) (both leaves of cherry 1 and one leaf each of cherries 2 and 3). Its three pair sums are \(8,9,9\). The two largest are equal, but
the gap to the third is \(1\), which is odd, whereas in a binary tree the gap is \(2m\ge2\). So (10) fails. In all, 216 quartets of \(\tau\) violate (10).

The positive result for \(n\le11\) is an exhaustive enumeration of all integer matrices satisfying (4)--(9), and every one of them satisfies (10); uniqueness at \(n=12\) combines a further enumeration with a structural argument showing that every counterexample is a 6-cherry lift of a matrix in a single \(S_6\)-orbit (Appendix C). The counterexample lies on the manifold, satisfies Kraft and the strong triangle, and is not a tree, so the convex manifold together with these linear conditions cannot certify that a matrix comes from a tree; other characterizations are not ruled out. Catanzaro, Pesenti, Sapucaia and Wolsey {[}13, Sec. 9{]} exhibit a \(12\times12\) matrix satisfying (4)--(9) but not (10), and it is isomorphic to ours, so the existence of a counterexample is due to them. Our additions are the threshold (\(n\le11\) sufficient), the uniqueness at \(n=12\), and the cherry-lift construction, which explains the odd gap.

This concludes the theoretical results. Part II turns to computation: it describes the solver built on them and compares it with the distance-indexed solver.

\part*{Part II: Computational results}

\section{7. Experimental setup}\label{experimental-setup}

In this section we describe the machine, the two solvers, the instances and the measures used in Section 8. Both solvers maintain an upper bound and a lower bound on the optimum: every tree they find gives an upper bound, its BME value, and the difficulty lies in raising the lower bound to match it. A solver \textbf{certifies} an instance when it closes the gap between the two, which proves that its best tree is optimal. The main text states each comparison's conclusion at the level of categories, such as whether a solver certifies or which cut family determines the bound; Appendix D gives the exact times, node counts and percentages behind each conclusion, and Appendix E the checks on numerical reliability.

\subsection{7.1 Hardware and software}\label{hardware-and-software}

\begin{itemize}
\tightlist
\item
  \textbf{Machine:} one Apple-silicon (arm64) laptop with 14 cores and 24 GB of RAM, running macOS. No two experiments that report a time ran at the same time.
\item
  \textbf{Threads:} every run that reports a time uses ten cores and a time limit of 3600 s: the distance-indexed solver runs Gurobi on 10 threads, and the compact solver runs ten worker processes of one thread each (Section 7.2). Nothing else runs beside it. Every other experiment reports a bound, a count or a check, uses one thread, and runs beside up to nine others.
\item
  \textbf{Random seeds:} a run that reports a time is repeated with three random seeds (Gurobi's \texttt{Seed} parameter; CPLEX's \texttt{randomseed} in the compact solver) when the first seed certifies, and we report the median and range of its times and node counts over the seeds. The distance-indexed solver is deterministic for a given seed and thread count, except where a run stops at the time limit; the compact solver with ten workers is not, since its node order depends on timing.
\item
  \textbf{Compact solver:} Python 3, with LPs solved by IBM CPLEX 22.1.2 through its Python API. A cut is separated when the LP point violates it by more than a separation tolerance \(\epsilon_{\mathrm{sep}}\). A node is pruned, and the root certifies, when its lower bound is at least the incumbent minus an absolute pruning tolerance \(\epsilon_{\mathrm{prune}}\). CPLEX's feasibility and optimality tolerances \(\epsilon_{\mathrm{LP}}\) are set below \(\epsilon_{\mathrm{sep}}\), so that the next LP point satisfies every separated cut. Appendix E gives the values of these and the other tolerances.
\item
  \textbf{Distance-indexed solver:} the C++ implementation of Catanzaro \emph{et al.} {[}13{]}, compiled with Apple clang 16 at \texttt{-O3} and linked to Gurobi 12.0.1. We ran the model of Section 2.4, whose quartet inequalities the implementation marks as lazy constraints (Gurobi's \texttt{Lazy} attribute \(1\)), so that Gurobi adds them to the relaxation when they cut off a candidate solution. We enabled every separation family at the implementation's default violation thresholds: circular order with a greedy tour, generalized Kraft, weak four-point and 2-K-split. The implementation sets Gurobi's relative MIP gap and keeps Gurobi's default tolerances (Appendix E). We call a run certified when its final relative gap \((UB-LB)/UB\) (Section 7.4) is \(0.0000\%\) to four decimals, that is, below \(5\times10^{-7}\). In both solvers, then, ``certified'' means optimal up to the solvers' floating-point tolerances; Appendix E checks the certified optima in exact arithmetic.
\item
  \textbf{Conic solver:} MOSEK 11.2.5 {[}30{]}, an interior-point solver for conic programs, run on one thread: directly in the compact solver's branch and bound, and through CVXPY 1.9.3 {[}31{]} in the root experiments (Section 7.2).
\item
  \textbf{Starting tree:} both solvers use FastME 2.1.6 {[}32{]}, which searches for a BME tree by local moves. The distance-indexed solver starts FastME from the tree built by its own LP-based variant of neighbour joining (NJ); the compact solver starts from the better of FastME's tree and the tree obtained from NJ by hill-climbing with NNI moves as steps.
\end{itemize}

\subsection{7.2 The compact solver}\label{the-compact-solver}

The compact solver runs its own branch and bound, with CPLEX solving the LP at each node and every separated cut kept for all later nodes. The search is \textbf{best-first}: it always processes next the open node with the least lower bound. A node commits some splits (\(W[S]=\tfrac12\)) and forbids others (\(W[S]\ge\tfrac34\)), a disjunction that is valid by Theorem 2. It branches on the most balanced split \(S\) of the incumbent (the best tree found so far) with \(\tfrac12<W[S]<\tfrac34\) at the LP optimum, into ``commit \(S\)'' and ``forbid \(S\)''; a split that crosses a committed one is only forbidden. A node is closed when its bound reaches the incumbent within the pruning tolerance \(\epsilon_{\mathrm{prune}}\), or when its LP optimum is a tree. When a child's relaxation still contains its parent's LP optimum, the child inherits the parent's bound, and its LP is solved only when the search selects it.

Merging two clusters of leaves whenever their union is a side of a committed split, from single leaves onwards, leaves clusters that each hang from the rest of the tree as a subtree of fixed shape (Lemma F.1). If leaf \(i\) lies \(e_i\) edges below the top of its cluster \(Z\), then in every tree of the node \(2^{e_i}w_{ic}\) is the same for every \(i\in Z\) and every leaf \(c\notin Z\), and \(w_{ij}=2^{-p_{ij}}\) for two leaves of \(Z\) at the fixed path length \(p_{ij}\). Each node imposes these \textbf{cluster equalities}, which its relaxation does not imply. When the LP optimum is no tree and has no split with \(\tfrac12<W[S]<\tfrac34\), the search enumerates the trees on the clusters if there are at most \(m_{\mathrm{enum}}\) of them. Otherwise it branches on the most balanced undecided half-cut (\(W[S]=\tfrac12\)) that crosses no committed split, and if there is none, on cherries: in every tree of the node some pair of the clusters attached to one end vertex of the tree that the
committed splits form on the clusters is a cherry, so child \(k\) commits the \(k\)-th such pair and forbids the earlier ones. Half-cuts are minimum cuts of the graph weighted by \(w\), since every cut is at least \(\tfrac12\), and the search lists them in polynomial time {[}33{]}. At such a node the search also looks for a better incumbent: it treats the cherries of the LP point as committed splits and enumerates the trees that remain. Appendix F states the branch and bound in pseudocode.

The search runs on \(k\) worker processes, each with its own copy of the relaxation. A coordinator holds the open nodes and the incumbent, and whenever a worker returns a node it passes that worker's cuts to the others and hands it the open node of least bound (or the worker's own unsolved child, if it has one). With \(k=1\) this is the sequential search.

We also implemented the families F4--F7 of Table 3 and inequalities (50), (53) and (54) of {[}12{]}; the \textbf{default cut set} omits them and keeps min-cut, manifold and PM, together with the \textbf{crossing cut} \(W[A]+W[B]\ge\tfrac54\) for two crossing bipartitions \(A\) and \(B\) that both have cut value \(\tfrac12\) at the LP optimum (valid by Theorem 2, since at most one of two crossing bipartitions is a split). We revisit this choice in Section 8.3.

\textbf{Imposing the manifold constraint.} The manifold constraint \(g(w)=\sum_{i<j}w_{ij}\log_2w_{ij}\le\tfrac32-n\) of the compact model is convex but not linear. We impose it exactly, as \(\sum_{i<j}\mathrm{entr}(w_{ij})\ge(n-\tfrac32)\ln2\) with \(\mathrm{entr}(x)=-x\ln x\), which an exponential-cone solver accepts directly. (We first imposed it in CPLEX by Kelley's tangent cuts {[}24{]}, but these led to numerical difficulties, and we dropped them.)

\begin{itemize}
\tightlist
\item
  \textbf{In the root experiments} (Sections 8.3, 8.4 and 8.7, Appendix E), every relaxation is solved by MOSEK, and the other cut families are separated at its optimum until none is violated.
\item
  \textbf{In the branch and bound}, each node's LP first goes through the cut loop in CPLEX, without the manifold constraint; the same rows with the constraint are then solved by MOSEK, and a cut violated at the conic optimum returns to the LP loop (Algorithm 5 of Appendix F). The node's bound is the larger of the LP bound and the conic bound below. Before each conic solve, every row is divided by the power of two nearest its largest coefficient (exact in floating point), since coefficients reach \(2^{n-4}\).
\end{itemize}

Each conic bound we report is rigorous. For multipliers \(y\) of the linear rows (with the sign each row's sense requires) and \(\mu\ge0\) of the manifold constraint, \[b^\top y-\mu\beta+\sum_j\min_{l_j\le w_j\le u_j}\bigl(r_jw_j+\mu\,w_j\ln w_j\bigr),
\qquad r=c-A^\top y,\quad\beta=(\tfrac32-n)\ln2,\] is a lower bound on the relaxation, whatever the multipliers; each one-dimensional minimum is at \(w_j=e^{-r_j/\mu-1}\) clipped to \([l_j,u_j]\). We take \(y\) and \(\mu\) from the conic solver's dual solution. In the root experiments we evaluate the bound in 60-digit arithmetic. In the branch and bound, which evaluates it at every node, we evaluate it in double precision and subtract a bound on every rounding error, so the result is at most the exact value (Appendix E). An inaccurate dual can only weaken this bound, never invalidate it. A relaxation with no nonlinear constraint (in Section 8.7, those that contain the distance-indexed model's linear manifold equality, which implies our convex inequality) is a linear program, and we solve it by the dual simplex method of CPLEX, with the safe bound of Neumaier and Shcherbina {[}34{]} described in Appendix E.

\subsection{7.3 Instances}\label{instances}

The instances come from two sources. (a) A synthetic testbed of integer distance matrices for \(n=4,\dots,10\), generated with varying distortion from a tree. They are nested, each being the leading principal submatrix of the next, so they are not independent instances. (b) The matrices distributed with the distance-indexed solver: Primates12, M17, M18 and M43, plus 42 supplementary matrices with \(n=20,\dots,30\), which we name by size and dataset as in that distribution (for example 20\_euros2). For the sweep of Section 8.6 we ran one supplementary matrix per size, chosen so that the same biological families continue across sizes (euros2 at \(n=20,22,23\); proteic M2624 at \(25,26\); nucleic M2573 at \(27,28\); B-NS1 at \(29,30\)). 20\_rosids and 20\_B-HA-573-585 come from the same supplementary set.

Objective values follow the scaling of the distance-indexed solver, whose value is ours times \(2^{n+1}\); for example, Primates12 is \(802.589=0.097972\times2^{13}\).

\subsection{7.4 Metrics}\label{metrics}

With certified optimum \(L^*\), lower bound \(LB\) and incumbent value \(UB\) (the BME value of the best tree found):

\begin{itemize}
\tightlist
\item
  \textbf{dual gap} \((L^*-LB)/L^*\); the \textbf{root gap} is the dual gap of the root LP after separation converges;
\item
  \textbf{primal gap} \((UB-L^*)/L^*\);
\item
  \textbf{solver gap} \((UB-LB)/UB\), as reported by Gurobi;
\item
  \textbf{gap closed by a family} \(F\): \((LB_{\text{base}+F}-LB_{\text{base}})/(L^*-LB_{\text{base}})\);
\item
  \textbf{root cut-closeable gap} \(LB_{\text{full}}-LB_{\text{base}}\), the bound that the full cut set adds to the base relaxation (Kraft and box). A family's \textbf{standalone} share is \((LB_{\text{base}+F}-LB_{\text{base}})/(LB_{\text{full}}-LB_{\text{base}})\), and its \textbf{marginal} share is \((LB_{\text{full}}-LB_{\text{full}-F})/(LB_{\text{full}}-LB_{\text{base}})\). The families overlap, so neither the standalone nor the marginal shares need sum to \(100\%\). These shares do not depend on the order in which the families are separated: when separation converges, the relaxation is the initial polytope with every inequality of the enabled families added, explicitly or implicitly, so it is well defined whatever the order. The order affects only how long convergence takes.
\end{itemize}

\section{8. Results}\label{results}

\subsection{\texorpdfstring{8.1 The compact solver certifies up to \(n=20\) and is mostly slower than the distance-indexed solver}{8.1 The compact solver certifies up to n=20 and is mostly slower than the distance-indexed solver}}\label{the-compact-solver-certifies-up-to-n20-and-is-mostly-slower-than-the-distance-indexed-solver}

In the timed solves (ten cores, Section 7.1), the compact solver certifies the optimum of every synthetic instance, of Primates12, M17 and M18, and of 20\_euros2 and 20\_B-HA, in every seed (Table D1). It certifies 20\_rosids in one of three seeds, just within the hour, and does not certify 21\_nucleic or 25\_proteic; every run it does not certify stops within \(1\%\) of the optimum. The distance-indexed solver, on the same ten cores, certifies all of them (Table D2). It is faster on Primates12, M18, 20\_euros2 and 20\_rosids; the compact solver is faster on M17 and 20\_B-HA, by factors of about \(1.3\) and \(2.3\) in median time, with the ranges over seeds disjoint. The compact solver's root gaps are larger, by factors between about 4 and 80 on M17, M18, 20\_euros2 and 20\_rosids (Tables D1 and D9), and its search makes up the difference by branching, with between five and nearly a thousand times as many nodes.

\subsection{8.2 The named facets close a shrinking share of the gap}\label{the-named-facets-close-a-shrinking-share-of-the-gap}

On the synthetic instances, the facet families F1 and F6 of Section 4, added to a base of Kraft, the box, every cut \(W[S]\ge\tfrac12\), the double cherry and every crossing cut \(W[A]+W[B]\ge\tfrac54\) of Section 7.2, close two thirds of the root gap at \(n=6\) and a tenth at \(n=10\) (Table D3).

\subsection{8.3 The minimum-cut family accounts for most of the root bound}\label{the-minimum-cut-family-accounts-for-most-of-the-root-bound}

For this attribution the manifold is one exact convex constraint (Section 7.2).

On the real instances, the min-cut family alone closes most of the gap that the full cut set closes at the root, between \(71\%\) and \(93\%\), with only 3 to 9 of its rows needed (Table D4). The manifold constraint closes nearly all the rest. Every other family (PM, F4--F7 and inequalities (50), (53) and (54) of {[}12{]}) adds at most a few percent over min-cut and nothing at the margin. This is why the default cut set keeps only min-cut, crossing, PM and manifold. The synthetic instance behaves in the opposite way: there the min-cut family closes almost nothing and PM most of the gap, so the synthetic testbed of Section 8.2 is not representative of real data.

\subsection{8.4 The split layer does not raise the root bound}\label{the-split-layer-does-not-raise-the-root-bound}

On Primates12, the membership model of Section 5.2 starts from a large Kraft-only gap; the min-cut family removes most of it and the manifold constraint most of the remainder. The split, coupling and laminarity layers then leave the root bound unchanged, with either laminarity encoding (Table D5). The coupling \(\sum_{i<j}P_{ijk}=\tfrac12\) pins \(w\) only at integral \(B\), and the pairwise laminarity constraints are big-M disjunctions, switched on or off by a binary variable through a large constant, which are loose at fractional \(s\). A further lifted model, with cut-value-ladder indicators (\(y_S=1\Rightarrow W[S]=\tfrac12\), \(y_S=0\Rightarrow W[S]\ge\tfrac34\)), the clique cover of Proposition 4 and an indicator for the \(\tfrac34\) rung, lowers the root gap below that of the min-cut relaxation but does not close it, consistent with \(G\) not being perfect (Section 5.1). Each of its three layers contributes, and the manifold is needed for the lowest gap.

The per-split cut-polytope triangles of Section 5.3 imply the aggregated strong triangle inequality and are strictly stronger as constraints on \(B\). Added to the level indicators of the distance-indexed model together with our \(B\) variables, they give exactly the same root bound as the strong triangle inequality alone, with or without the min-cut family (Table D6); we did not investigate why any further. The odd-cycle inequalities of the cut polytope are never violated at the root optimum of the level-indicator model. In the membership model they are violated, but adding them does not change the root bound.

\subsection{8.5 The residual cannot be repaired quartet by quartet}\label{the-residual-cannot-be-repaired-quartet-by-quartet}

At the root optimum \(w^*\) of the compact model, with the manifold imposed exactly, on M17, M18, 20\_euros2 and 20\_rosids, we call a quartet \textbf{violated} if \(w^*\) violates one of the strong four-point inequalities \(w_{ab}w_{cd}\ge4w_{ac}w_{bd}\) and \(w_{ab}w_{cd}\ge4w_{ad}w_{bc}\) of the quartet's resolution \(ab|cd\) in the certified optimal tree \(T\). Between a third and a half of all quartets are violated, and among the more strongly violated half \(w^*\) mostly prefers a resolution other than \(T\)'s (Table D7). Since this uses \(T\), it is a diagnostic and not a separation procedure.

Imposing \(T\)'s quartets globally, as the distance-indexed model does lazily, needs a solver that handles both the non-convex four-point products and the exponential cone of the manifold. MOSEK solves the cone but not the products. Gurobi handles the products by spatial branch and bound but has no exponential cone, so it must outer-approximate \(w\ln w\); on M17 its bound stayed below the convex root's even before any quartet was added. We therefore ask a narrower question: whether the violations can be fixed locally.

To see whether they can, we try to repair each violated quartet on its own, changing only its six weights. The changes that keep Kraft form a plane, \[\Delta_{ab}=\Delta_{cd}=s,\qquad\Delta_{ac}=\Delta_{bd}=t,\qquad\Delta_{ad}=\Delta_{bc}=-(s+t),\] since the Kraft rows of the quartet's four leaves have rank four on its six pairs. A \textbf{local fix} is a point \((s,t)\) at which the quartet satisfies \(T\)'s two inequalities. (We do not impose the equality \(w_{ac}w_{bd}=w_{ad}w_{bc}\) of Section 6.1, which can only make a fix harder.) We write \(P_0=w_{ab}w_{cd}\), \(P_1=w_{ac}w_{bd}\) and \(P_2=w_{ad}w_{bc}\). On the plane, the manifold function \(g(w)=\sum_{i<j}w_{ij}\ln w_{ij}\) has gradient \((\ln(P_0/P_2),\ln(P_1/P_2))\), so it is smallest where the three products are equal and grows as they separate, while the four-point condition asks for \(P_0\ge4\max(P_1,P_2)\). Because \(w^*\) lies on the boundary of the manifold constraint, a fix that keeps the manifold must stay within
the level set of \(g\) through \(w^*\). For each violated quartet we compute, with Gurobi's global solver on the two variables, the local fix of least manifold excess, first within the box \(2^{-(n-1)}\le w\le\tfrac14\) only and then with every row of the root relaxation imposed as well.

Within the box, almost every violated quartet has a local fix, and for more than half of them the fix keeps the manifold constraint; nine in ten of the others need an excess \(g\) above its bound of under \(0.003\) bits, and none more than \(0.04\). The manifold is therefore not what ties the quartets together. The cut rows are: once the root's other rows are imposed, between \(85\%\) and \(97\%\) of the violated quartets have no local fix at all. Each violated quartet's six weights appear in a median of 6 to 14 rows that are tight at \(w^*\) and that any local fix would change. The share is as high among the quartets whose separation \(\log_2P_0-\log_2\max(P_1,P_2)\) is below \(1.75\), that is, falls short of the required \(2\) by more than \(12.5\%\). At the other extreme, the optimal tree's point, which satisfies every quartet, differs from \(w^*\) in nearly all of the \(\binom n2\) weights (Table D7). The four-point residual at the root is thus a property of the whole weight
vector: it cannot be repaired one quartet at a time without breaking the cuts that hold the root bound.

\subsection{\texorpdfstring{8.6 The distance-indexed model certifies 13 of 14 instances with \(n\le30\) in every seed}{8.6 The distance-indexed model certifies 13 of 14 instances with n\textbackslash le30 in every seed}}\label{the-distance-indexed-model-certifies-13-of-14-instances-with-nle30-in-every-seed}

In the timed solves (10 threads), the distance-indexed solver certifies 13 of the 14 instances with \(12\le n\le30\) in all three seeds (Table D8). The fourteenth, 28\_nucleic, certifies in two seeds and stops at the time limit with a gap of \(0.05\%\) in the third. Times vary between seeds by up to a factor of four on the same instance, so they support only category-level comparisons. On M43 (\(n=43\)) the solver does not leave the root node within the hour, and stops with a gap of \(0.9\%\). M43 is also the only instance at which the smallest terms \(D_{ij}2^{-\tau_{ij}}\) of the objective approach machine precision relative to the objective, the regime in which Catanzaro, Frohn and Pesenti {[}35{]} show the problem becomes numerically unstable. Every certified run of an instance returns the same tree (Appendix E), which we recover from the integer \(\tau\) output by \textbf{cherry contraction}, repeatedly replacing a cherry by a single leaf; Section 8.5 uses these optima.

\subsection{8.7 Its root bound comes from its integer lifting, and four-point separation makes its certificates correct}\label{its-root-bound-comes-from-its-integer-lifting-and-four-point-separation-makes-its-certificates-correct}

To separate the formulation from its cut families, we measured the root gap to \(L^*\) with the families switched on cumulatively (Table D9): (a) the base model with no separated cuts, (b) strong triangles on all triples, then generalized Kraft, then 2-K-split and circular order, then four-point. Configurations (a) and (b) are root LPs; the others are full solves, of which we report the root bound.

\begin{itemize}
\tightlist
\item
  The base model already contains every triangle inequality \(\tau_{ik}+\tau_{kj}-\tau_{ij}\ge2\) and the manifold as a linear equality in the level indicators (Section 6.2). Strengthening the triangles by parity, as in (b), does not move the root LP bound.
\item
  With generalized Kraft (our min-cut family) separated, the root bound of the full solve closes between \(71\%\) and \(98\%\) of the base gap, and the 2-K-split and circular-order families add nothing further at the root. That bound, however, is not the bound of the model's linear relaxation: the MIP solver also adds its own general-purpose cutting planes, which use the integrality of the level indicators and of the parity variables. With those switched off and the solve stopped at the root (Table D11), generalized Kraft closes between \(2\%\) (M18) and \(72\%\) (20\_rosids) of the base gap, and Gurobi's cuts close most of the rest; on M17 they are mainly mixed-integer rounding, flow cover, cover and GUB cover cuts.
\item
  Without four-point separation the model is a relaxation of BME. With generalized Kraft as the only separated family, on M18 and 20\_rosids it ends at a matrix that satisfies (4)--(9) of Section 6.3 and has cherries, but is not a tree, with a value below \(L^*\). These are real-data instances of the phenomenon behind Theorem 5. With the 2-K-split and circular-order families added, the M18 run certifies \(L^*\) and the 20\_rosids run does not finish within the hour; neither ends at a non-tree.
\item
  Four-point separation makes certification much faster, as measured by the number of nodes: from \(71\) to \(1\) on M17 and from \(5{,}263\) to \(965\) on M18, while 20\_rosids, which does not certify within the hour without it, certifies in \(131\) nodes (medians over the seeds). It does so during the search: imposed statically on the quartets containing leaf 1, the four-point rows leave the root bound unchanged (Table D10), and the solver adds them lazily, where they cut off the non-tree points the search reaches.
\end{itemize}

How much of the root comes from general-purpose cuts depends on the solver and the instance. Given the same static model (without generalized Kraft), Gurobi's and CPLEX's cuts at the root close between \(12\%\) and \(85\%\) of its LP gap, and neither solver is better on every instance (Table D12). On 28\_nucleic they close only about a quarter of it, and the separated families do most of the work: its root falls from about \(2.1\%\) to \(0.3\%\) once they are added.

Which rows of the distance-indexed model would help the compact one can be tested by adding them to the compact relaxation (Table D10). The level indicators alone, the triangles alone, the solver's root 2-K-split separation and the leaf-1 four-point rows each leave the compact root bound unchanged. The linear manifold equality closes between \(6\%\) and \(47\%\) of the compact root gap, and together with the triangles between \(23\%\) and \(90\%\); the lifted 2-K-splits add at most \(5\%\). With all of these rows, the compact relaxation is at least as tight as the distance-indexed model's own LP with generalized Kraft, on every instance tested (Table D11).

The distance-indexed model's advantage at the root therefore lies in no particular cut family, but in its lifting to integer level indicators. The lift makes the whole manifold equality linear, where the compact model can impose only its convex half, and it gives the MIP solver integer variables on which its general-purpose cuts tighten the root. Four-point separation during the search then makes the certificates correct and fast.

The two solvers also differ in presolve, branching, heuristics and implementation language. These measurements isolate the cut families and the base rows, but they do not separate the formulation from the implementation.

\section{9. Discussion and open problems}\label{discussion-and-open-problems}

The weight-space formulation has much more structure than earlier work used. A fixed tree's weights are determined by linear split data and one convex constraint (Theorem 1), cut values form a discrete ladder with an exact NNI characterization and a topology-invariant count (Theorems 2 and 3), compatibility has a polynomial clique description (Proposition 4), and the classical path-length conditions without the four-point condition suffice exactly up to \(n=11\) (Theorem 5). These results stand independently of any solver.

Computationally, the compact model uses \(\binom n2\) weight variables against \(\Theta(n^3)\) for the distance-indexed model, keeps the objective linear, and certifies instances with up to \(n=20\) taxa within an hour. The distance-indexed model certifies most of them faster, from a tighter root bound, and certifies 13 of 14 instances with \(n\le30\) in every seed. The measurements place the difference in the distance-indexed model's integer lifting: the linear manifold equality it allows, and above all the integer level indicators on which a MIP solver's general-purpose cuts act; four-point separation then makes its certificates correct and fast. They do not show that no compact model, and no better branching or separation, could close the gap.

\textbf{Open problems.} Each item is labelled as a question, a conjecture or an observation.

\begin{itemize}
\tightlist
\item
  \textbf{Extension complexity in \(w\) (question).} An extended formulation of a polytope is a polytope in more variables that projects onto it, and the extension complexity is the smallest number of facets such a polytope can have. Is there an extended formulation of \(P_{\mathrm{BME}}\), in \(w\) or in split indicators, with \(o(n^3)\) variables that represents the manifold equality linearly and exposes integer structure that general-purpose MIP cuts can exploit? Our results bear only on unlifted \(w\)-constraints. As far as we know, lower-bound techniques such as those of {[}36{]}, {[}37{]} have not been applied to \(P_{\mathrm{BME}}\).
\item
  \textbf{Conjecture 2 (a cherry always exists).} Every matrix satisfying (4)--(9) has a cherry, that is, \(\tau_{ij}=2\) for some pair. It holds by exhaustive search for \(n\le15\), and constraint programming (\(n=16\)--\(20\)) and simulated annealing (\(n=24\)) found no counterexample.
\item
  \textbf{The \(n\le11\) coincidence (observation).} The sufficiency threshold of Theorem 5 coincides with the range \(n\le11\) in which the splitohedron, an outer relaxation of \(P_{\mathrm{BME}}\) defined in {[}10{]}, equals the BME polytope {[}14, p. 17{]}. The two objects are different, and we do not claim a common cause.
\end{itemize}

In summary, the weight space describes a tree exactly through linear split data and one convex constraint, and a compact solver built on this description certifies instances with up to \(20\) taxa. Its weakness is the root bound: the distance-indexed model's integer lifting, which the compact model lacks, is what lets that solver certify larger instances. Closing this gap without giving up the compactness of the weight space is the main open problem that our results leave.

\section{Appendix A. The star-point inequality}\label{appendix-a.-the-star-point-inequality}

We show that \(g(w^\star)<\tfrac32-n\) for every integer \(n\ge4\), where \(w^\star_{ij}=\frac1{2(n-1)}\). There are \(\binom n2\) pairs, so \[g(w^\star)=\binom n2\frac1{2(n-1)}\log_2\frac1{2(n-1)}=-\frac n4\bigl(1+\log_2(n-1)\bigr).\] The inequality \(-\frac n4(1+\log_2(n-1))<\tfrac32-n\) is equivalent, after dividing by \(n/4\) and moving the \(n\)-dependent terms to the left, to \[f(n):=\log_2(n-1)+\frac6n>3 .\] For real \(x>1\), \[f'(x)=\frac1{(x-1)\ln2}-\frac6{x^2}=\frac{q(x)}{x^2(x-1)\ln2},\qquad q(x)=x^2-6\ln2\,(x-1).\] Since \(q'(x)=2x-6\ln2>0\) for \(x>3\ln2\), and \(3\ln2<3\), the quadratic \(q\) is increasing on \([3,\infty)\). Also \(q(3)=9-12\ln2>0\) because \(\ln2<\tfrac34\). So \(q>0\) and \(f\) is strictly increasing on \([3,\infty)\). The minimum of \(f\) on \((1,\infty)\) lies at the larger root of \(q\), between \(2\) and \(3\), and at the neighbouring integers \(f(2)=f(3)=3\). Hence \(f(n)>3\) for every integer \(n\ge4\). (At \(n=3\) the star is the only binary
tree, and the inequality becomes an equality.) \(\square\)

\section{Appendix B. Proofs of Theorems 2 and 3}\label{appendix-b.-proofs-of-theorems-2-and-3}

Throughout, \(T\) is binary, \(S\) is a bipartition, and leaves are coloured \(S\) or \(S^c\).

\textbf{Mass fact.} If \(X\) is the component of \(T-uv\) containing \(v\), then \(\sum_{a\in X}2^{-d(u,a)}=\tfrac12\). \emph{Proof:} rooted at \(v\), \(X\) is a full binary tree (or the single leaf \(v\)), so \(\sum_{a\in X}2^{-d(v,a)}=1\) by the Kraft equality for prefix codes; the edge \(uv\) adds a factor \(\tfrac12\).

\textbf{Lemma B.1 (structure of the boundary).} (a) Pendant edges are never boundary edges. (b) \(B(S)\) is connected. (c) For a vertex \(u\) of \(B(S)\), every component of \(T-u\) that contains no boundary edge is monochromatic. (d) A vertex of degree \(1\) in \(B(S)\) has two pendant subtrees of opposite colours. (e) Conversely, let \(\pi\) be the path between two distinct internal vertices, and colour the pendant subtrees hanging off \(\pi\) so that each end of \(\pi\) has both colours. Then the induced \(S\) has \(B(S)=\pi\).

\emph{Proof.} (a) One side is a single leaf. (b) Let \(e,f\in B(S)\) and let \(h\) lie on the path between them. Each side of \(h\) contains the far side of \(e\) or of \(f\), which is bichromatic, so \(h\in B(S)\). (c) Let \(uv\) be the edge into such a component \(X\). Since \(uv\notin B(S)\), one side of \(uv\) is monochromatic. The \(u\)-side contains the far side of a boundary edge at \(u\), which is bichromatic, so \(X\) is monochromatic. (d) The side of the unique boundary edge at \(u\) that contains \(u\) is the union of \(u\)'s two pendant subtrees. It is bichromatic, and each pendant subtree is monochromatic by (c). (e) Each edge of \(\pi\) has both colours on each side, coming from the end of \(\pi\) on that side. Each edge off \(\pi\) has a side inside one pendant subtree. \(\square\)

\textbf{Lemma B.2 (empty boundary).} \(B(S)=\emptyset\) if and only if \(S\) is a split, and then \(W_T[S]=\tfrac12\).

\emph{Proof.} If \(S\) is the side of an edge \(uv\) containing \(u\), then every edge has a side inside \(S\) or inside \(S^c\), so \(B(S)=\emptyset\). Applying the mass fact on both sides gives \(W_T[S]=\sum_{a\in S,b\notin S}2^{-d(a,u)-1-d(v,b)}=1\cdot1\cdot\tfrac12\). Conversely, suppose \(B(S)=\emptyset\) but \(S\) is not a split. Among all monochromatic sides of edges, we take one, \(M\), of maximum size, and suppose that \(M\) is the \(v\)-side of \(uv\) and has colour \(S\). Then \(M\ne S\), so the \(u\)-side contains leaves of both colours and \(u\) is internal, with two further edges \(f,g\) whose far sides \(F,G\) partition the leaves on the \(u\)-side. The near side of \(f\) contains \(M\cup G\), which is larger than \(M\), so by maximality it is not monochromatic. Hence \(F\) is monochromatic, and likewise \(G\). Since \(F\cup G\) contains both colours, one of them, say \(F\), has colour \(S\). Then \(M\cup F\) is a monochromatic side of \(g\) larger than \(M\), a
contradiction. \(\square\)

\textbf{Lemma B.3 (path identity).} Let \(\ell_1,\ell_2\) be distinct vertices of degree \(1\) in \(B(S)\ne\emptyset\), joined by the path \(v_0=\ell_1,v_1,\dots,v_k=\ell_2\) (\(k\ge1\)). Each interior \(v_i\) has one neighbour off the path; let \(X_i\) be the component through it. We define \(m_S(i)=2\sum_{a\in X_i\cap S}2^{-d(v_i,a)}\) and \(m_c(i)=1-m_S(i)\), the two colour masses of \(X_i\), which sum to \(1\) by the mass fact. Then \[4W_T[S]=4-2^{1-k}+\sum_{1\le i<j\le k-1}\bigl(m_S(i)m_c(j)+m_c(i)m_S(j)\bigr)2^{-(j-i)}+4\sum_{i=1}^{k-1}W_{X_i},\] where \(W_{X_i}=\sum_{a\in X_i\cap S,\,b\in X_i\cap S^c}2^{-\tau_{ab}}\ge0\).

\emph{Proof.} Every leaf lies either in one of the four end pendant subtrees (two at each end, of opposite colours by B.1(d)) or in some \(X_i\). By the mass fact, a part of mass \(\alpha\) at path position \(i\) and a part of mass \(\beta\) at position \(j\) contribute \(\alpha\beta\,2^{-|i-j|}\) to \(4W\), and each end pendant subtree has mass \(1\). There are four kinds of opposite-coloured pairs.

\begin{itemize}
\tightlist
\item
  End--end: \(1+1\) (same vertex) plus \(2\cdot2^{-k}\) (across).
\item
  End--\(X_i\): each colour part of \(X_i\) meets the opposite-coloured pendant subtree at each end, giving \((m_S(i)+m_c(i))(2^{-i}+2^{-(k-i)})=2^{-i}+2^{-(k-i)}\). Summed over \(i=1,\dots,k-1\) this is \(2-2^{2-k}\).
\item
  \(X_i\)--\(X_j\): the displayed double sum.
\item
  Within \(X_i\): the last term.
\end{itemize}

Adding, \(2+2^{1-k}+2-2^{2-k}=4-2^{1-k}\). \(\square\)

\textbf{Proof of Theorem 2.} Part 1 is Lemma B.2.

\emph{Path boundaries.} If \(B(S)\) is the path \(v_0\cdots v_k\), each \(X_i\) is a pendant subtree, monochromatic by B.1(c), so \(m_S(i)\in\{0,1\}\) and \(W_{X_i}=0\). Lemma B.3 gives \(4W=4-2^{1-k}+P\). Hence \(W=1-2^{-(k+1)}\) if and only if the interior colouring is constant, which is always the case for \(k\le2\). For a non-constant colouring with \(k\ge3\), some adjacent interior pair differs, so \(P\ge\tfrac12>2^{1-k}\) and \(W>1\).

For \(k=1\), \(W=\tfrac34\). Let the boundary edge be \(uv\), with pendant subtrees \(A,B\) at \(u\) and \(C,D\) at \(v\). By B.1(d), \(S\) is \(A\cup C\) or \(A\cup D\) up to complement, and these are exactly the two splits created by the NNI moves across \(uv\). Conversely, each NNI split across \(uv\) has boundary \(\{uv\}\). The boundary determines the edge, so the \(2(n-3)\) bipartitions are distinct. Bipartitions with \(|B|\ge2\) have \(W\ge\tfrac78\), so \(W=\tfrac34\) if and only if \(|B(S)|=1\). This proves parts 2 and 3.

\emph{Branching boundaries.} If \(B(S)\) is not a path, it is a subtree with at least three vertices of degree \(1\). We take two of them, \(\ell_1\) and \(\ell_2\), at distance \(k\ge2\). A third, \(\ell_3\), lies off the path, inside some \(X_i\). Its two pendant subtrees have opposite colours and are at distance \(0\), so \(4W_{X_i}\ge1\). Lemma B.3 gives \(4W\ge4-2^{1-k}+1\ge\tfrac92\), that is, \(W\ge\tfrac98\). Equality holds for three cherries around one vertex with \(S\) taking one leaf of each (Section 3.3). This proves part 4. \(\square\)

\textbf{Proof of Theorem 3.} By Theorem 2, \(W\le1\) if and only if \(B(S)=\emptyset\) (\(2n-3\) splits, one per edge) or \(B(S)\) is a path with constant colouring. By B.1(e) and the uniqueness of \(B(S)\), bipartitions of the second kind correspond bijectively to pairs consisting of a path between distinct internal vertices and an admissible colouring up to complement. For a path of length \(k\) there are \(2\cdot2\cdot2^{k-1}\) admissible colourings, of which \(8\) are constant for \(k\ge2\); halving for complements leaves \(4\). For \(k=1\) there are \(4/2=2\). There are \(\binom{n-2}2\) pairs of internal vertices, \(n-3\) of them adjacent, so \[|\{W\le1\}|=(2n-3)+2(n-3)+4\Bigl[\tbinom{n-2}2-(n-3)\Bigr]=2(n-3)^2+2n-3=2n^2-10n+15.\ \square\]

\section{Appendix C. The enumeration behind Theorem 5}\label{appendix-c.-the-enumeration-behind-theorem-5}

\textbf{Representation.} Candidate matrices are filled row by row in the upper triangle. A \textbf{row type} is a multiset of path lengths that can appear in one Kraft row. Because \(\sum_j2^{-(\tau_{ij}-1)}=1\) is the depth profile of a full binary tree, there are few row types: 9, 16, 28, 50 and 89 for \(n=8,\dots,12\).

\textbf{Pruning.} The search applies three prunes.

\begin{enumerate}
\def\labelenumi{\arabic{enumi}.}
\tightlist
\item
  It checks the strong triangle (6) as soon as a triple is complete.
\item
  Every partially filled row must remain a sub-multiset of at least one row type, tracked as a bitmask over the types. This makes (7) and (8) exact at completion.
\item
  It bounds (9) row by row, using the minimum and maximum manifold contribution over each incomplete row's remaining compatible types.
\end{enumerate}

\textbf{Symmetry and completeness.} Leaf 1's row is put in non-decreasing order, which removes relabellings of the other \(n-1\) leaves. This is valid for an existence question, since any solution can be relabelled to satisfy it. Otherwise the search is exhaustive: it cuts a branch only if the branch violates (6)--(9). A \textbf{survivor} is a complete matrix satisfying (4)--(9), and each survivor is then tested against (10). Without the ordering constraint on leaf 1, the survivor counts equal \((2n-5)!!\) for \(n\le8\) with no non-tree, so the enumerator neither misses a tree nor produces a spurious matrix on these sizes.

\textbf{Table C1. Canonical survivors of the enumeration.} For \(n=12\) see ``Uniqueness at \(n=12\)'' below.

{\def\LTcaptype{none} 
\begin{longtable}[]{@{}llllllll@{}}
\toprule\noalign{}
\(n\) & 5 & 6 & 7 & 8 & 9 & 10 & 11 \\
\midrule\noalign{}
\endhead
\bottomrule\noalign{}
\endlastfoot
survivors & 4 & 7 & 19 & 85 & 502 & 1252 & 5008 \\
violating (10) & 0 & 0 & 0 & 0 & 0 & 0 & 0 \\
\end{longtable}
}

\textbf{Uniqueness at \(n=12\).} (o) A complete search for \(12\times12\) matrices satisfying (4)--(9) with no pair at path length 2 finds none, so every counterexample has a cherry. (i) Let \(\{i,j\}\) be a cherry of a counterexample \(\tau\). The strong triangle forces \(\tau_{ik}=\tau_{jk}=:a_k\) for all \(k\). Contracting the cherry to a leaf \(p\) with \(\tau_{pk}=a_k-1\) preserves Kraft and range, lowers the manifold sum by exactly \(1\) (so (9) holds for \(n-1\)), and preserves the strong triangle except possibly on triangles through \(p\). If it held there too, the contracted matrix would satisfy (4)--(9) for \(n=11\) and so would be a tree, and \(\tau\), its cherry expansion, would be a tree as well, a contradiction. So some \(k,l\) have \(a_k+a_l-\tau_{kl}<4\). In the quartet \(\{i,j,k,l\}\) the two cross sums then both equal \(a_k+a_l\) and the third is \(2+\tau_{kl}\), so the gap is below \(2\) (or the third sum is the unique maximum) and (10) fails. Every cherry of a
counterexample therefore lies in a violating quartet. (ii) Hence a complete search that fixes a cherry at \(\{1,2\}\) and a violation at \(\{1,2,3,4\}\) finds every class of counterexample. It terminates with 84 witnesses, all in one class: 6-cherry lifts. (iii) The reduced 6-node matrices of weak-triangle type number 285, of which 105 are trees and 180 are not, and the 180 form a single \(S_6\)-orbit (that of \(r\), whose stabilizer has order 4).

\section{Appendix D. Tables of results}\label{appendix-d.-tables-of-results}

The tables of this appendix give the exact values behind the conclusions of Section 8. Times are wall-clock seconds; a cell with a range is the median and, in brackets, the minimum and maximum over the seeds that ran.

\textbf{Table D1. The compact solver} (Section 8.1). Ten cores. Ten workers of one thread each, time limit 3600 s; seeds 2 and 3 where seed 1 certified. Time in seconds. Gaps relative to \(L^*\); the final gap is given for runs stopped by the time limit.

{\def\LTcaptype{none} 
\begin{longtable}[]{@{}llllll@{}}
\toprule\noalign{}
instance & certified & nodes & time (s) & root gap & final gap \\
\midrule\noalign{}
\endhead
\bottomrule\noalign{}
\endlastfoot
test\_n5 & 3 of 3 & 0 & 0.7 {[}0.7--0.8{]} & -- & -- \\
test\_n6 & 3 of 3 & 5 & 0.7 {[}0.7--0.8{]} & 2.45\% & -- \\
test\_n7 & 3 of 3 & 11 & 0.8 {[}0.8--0.9{]} & 2.30\% & -- \\
test\_n8 & 3 of 3 & 88 {[}87--88{]} & 0.8 {[}0.8--0.9{]} & 8.00\% & -- \\
test\_n9 & 3 of 3 & 287 {[}285--289{]} & 1.1 {[}1.1--1.2{]} & 9.86\% & -- \\
test\_n10 & 3 of 3 & 652 {[}648--656{]} & 2.0 {[}2.0--2.1{]} & 13.57\% & -- \\
Primates12 & 3 of 3 & 27 & 1.1 {[}1.1--1.2{]} & 0.64\% & -- \\
M17 & 3 of 3 & 457 & 5.8 {[}5.7--5.8{]} & 0.80\% & -- \\
M18 & 3 of 3 & 10361 {[}10359--10372{]} & 266 {[}255--289{]} & 2.61\% & -- \\
20\_euros2 & 3 of 3 & 867 {[}867--869{]} & 22.3 {[}21.1--23.8{]} & 0.56\% & -- \\
20\_B-HA & 3 of 3 & 317 & 10.4 {[}9.9--10.6{]} & 0.82\% & -- \\
20\_rosids & 1 of 3 & 42771 {[}42534--45724{]} & 3600 {[}3582--3601{]} & 1.07\% & 0.011\% {[}0.011\%--0.012\%{]} \\
21\_nucleic & 0 of 1 & 42437 & 3600 & 2.55\% & 0.915\% \\
25\_proteic & 0 of 1 & 15790 & 3601 & 1.27\% & 0.379\% \\
\end{longtable}
}

\textbf{Table D2. The two solvers on the instances of Table D1 with \(n\ge12\)} (Section 8.1). Ten cores. The compact solver as in Table D1, the distance-indexed solver in the configuration of Section 7.1 (Table D8; 20\_rosids from the configuration full of Table D9, which is the same). Time in seconds to certify, and nodes, for each solver.

{\def\LTcaptype{none} 
\begin{longtable}[]{@{}lllll@{}}
\toprule\noalign{}
instance & compact (s) & nodes & distance-indexed (s) & nodes \\
\midrule\noalign{}
\endhead
\bottomrule\noalign{}
\endlastfoot
Primates12 & 1.1 {[}1.1--1.2{]} & 27 & 0.6 & 1 \\
M17 & 5.8 {[}5.7--5.8{]} & 457 & 7.7 {[}6.8--8.9{]} & 1 \\
M18 & 266 {[}255--289{]} & 10361 {[}10359--10372{]} & 94.5 {[}78.4--103{]} & 2081 {[}1717--2164{]} \\
20\_euros2 & 22.3 {[}21.1--23.8{]} & 867 {[}867--869{]} & 6.5 {[}6.4--6.6{]} & 1 \\
20\_B-HA & 10.4 {[}9.9--10.6{]} & 317 & 23.9 {[}22.3--25.2{]} & 1 \\
20\_rosids & 1 of 3 certified & 42771 {[}42534--45724{]} & 61.2 {[}57.3--65.9{]} & 230 {[}127--376{]} \\
21\_nucleic & 0 of 1 certified & 42437 & 253 {[}189--269{]} & 597 {[}391--780{]} \\
25\_proteic & 0 of 1 certified & 15790 & 321 {[}285--435{]} & 852 {[}733--1314{]} \\
\end{longtable}
}

\textbf{Table D3. Root bound with F1 and F6 on the synthetic instances} (Section 8.2). One thread. \(L^*\) by enumeration of all trees.

{\def\LTcaptype{none} 
\begin{longtable}[]{@{}llllll@{}}
\toprule\noalign{}
\(n\) & \(L^*\) & base & \(+\)F1 & \(+\)F1\(+\)F6 & gap closed \\
\midrule\noalign{}
\endhead
\bottomrule\noalign{}
\endlastfoot
6 & 5.0938 & 4.6964 & 4.8125 & 4.9643 & 67\% \\
7 & 5.4375 & 5.0417 & 5.1272 & 5.1660 & 31\% \\
8 & 6.0156 & 5.2604 & 5.3581 & 5.3979 & 18\% \\
9 & 6.7109 & 5.6806 & 5.7646 & 5.8146 & 13\% \\
10 & 7.0195 & 5.7020 & 5.7774 & 5.8283 & 10\% \\
\end{longtable}
}

\textbf{Table D4. Shares of the root cut-closeable gap} (Section 8.3; definitions in Section 7.4). One thread. With the manifold imposed exactly (Section 7.2) and every other family separated by the solver's own separators until none is violated. Min-cut is standalone; the manifold and the other families are measured with min-cut on: ``incr.'' is the increment over min-cut alone and ``marg.'' the marginal share in the full set. ``Tight'' counts the min-cut rows tight at the root optimum. ``Other'' gives the other family with the largest increment, and the largest marginal share of any other family in absolute value.

{\def\LTcaptype{none} 
\begin{longtable}[]{@{}lllllll@{}}
\toprule\noalign{}
instance & min-cut & tight & manifold incr. & manifold marg. & other incr. & other marg. \\
\midrule\noalign{}
\endhead
\bottomrule\noalign{}
\endlastfoot
test\_n10 & 3\% & 1 of 1 & 48\% & 2\% & PM 74\% & 16.2\% \\
Primates12 & 93\% & 5 of 5 & 7\% & 4\% & PM 2\% & 0.0\% \\
M17 & 82\% & 3 of 3 & 18\% & 16\% & PM 2\% & 0.0\% \\
M18 & 71\% & 3 of 3 & 29\% & 26\% & PM 2\% & 0.0\% \\
20\_B-HA-573-585 & 89\% & 9 of 9 & 11\% & 10\% & PM 1\% & 0.0\% \\
\end{longtable}
}

\textbf{Table D5. Root gap on Primates12 of the membership model and the lifted ladder model} (Section 8.4). One thread. Left: membership model, layers added cumulatively, with the two laminarity encodings. Right: lifted ladder model.

{\def\LTcaptype{none} 
\begin{longtable}[]{@{}lll@{}}
\toprule\noalign{}
membership model & quadrant & nested/disjoint \\
\midrule\noalign{}
\endhead
\bottomrule\noalign{}
\endlastfoot
Kraft & 16.86\% & 16.86\% \\
+ min-cut & 4.81\% & 4.81\% \\
+ manifold & 1.56\% & 1.56\% \\
+ split layer and coupling & 1.56\% & 1.56\% \\
+ laminarity & 1.56\% & 1.56\% \\
\end{longtable}
}

{\def\LTcaptype{none} 
\begin{longtable}[]{@{}llll@{}}
\toprule\noalign{}
\(\tfrac34\)-rung indicator & cliques & manifold & root gap \\
\midrule\noalign{}
\endhead
\bottomrule\noalign{}
\endlastfoot
no & none & no & 3.42\% \\
no & none & yes & 1.56\% \\
no & Prop. 4 & no & 2.71\% \\
no & Prop. 4 & yes & 1.20\% \\
yes & none & no & 2.24\% \\
yes & none & yes & 1.41\% \\
yes & Prop. 4 & no & 1.57\% \\
yes & Prop. 4 & yes & 1.04\% \\
\end{longtable}
}

\textbf{Table D6. Root gap on Primates12 of the level-indicator model with per-split triangles} (Section 8.4). One thread. The last column is the largest difference between the bound with per-split triangles (alone or with the strong triangle) and the bound with the strong triangle alone.

{\def\LTcaptype{none} 
\begin{longtable}[]{@{}llllll@{}}
\toprule\noalign{}
min-cut & neither & strong triangle & per-split & both & difference \\
\midrule\noalign{}
\endhead
\bottomrule\noalign{}
\endlastfoot
off & 2.471\% & 2.071\% & 2.071\% & 2.071\% & 1e-17 \\
on & 1.001\% & 0.344\% & 0.344\% & 0.344\% & 1e-17 \\
\end{longtable}
}

\textbf{Table D7. Local fixes of the quartets violated at the compact model's root} (Section 8.5). One thread. At the root optimum \(w^*\) (manifold exact): quartets whose optimal-tree resolution \(w^*\) violates; the share of them with no local fix (a change of the quartet's six weights that keeps Kraft and satisfies the tree's four-point inequalities) within the box \(2^{-(n-1)}\le w\le\tfrac14\); with a local fix within the box that also keeps the manifold constraint; with no local fix once every row of the root relaxation is imposed as well (rows), over all violated quartets and over those violated by more than 12.5\% (separation below 1.75 bits); and the number of weights, out of \(\binom n2\), in which \(w^*\) differs from the optimal tree's point.

{\def\LTcaptype{none} 
\begin{longtable}[]{@{}llllllll@{}}
\toprule\noalign{}
instance & root gap & violated & no fix (box) & keeps manifold & no fix (rows) & \textgreater{} 12.5\% & differ \\
\midrule\noalign{}
\endhead
\bottomrule\noalign{}
\endlastfoot
M17 & 0.80\% & 963 (40\%) & 5\% & 54\% & 96\% & 97\% & 136/136 \\
M18 & 2.61\% & 1464 (48\%) & 3\% & 64\% & 85\% & 86\% & 152/153 \\
20\_euros2 & 0.56\% & 1682 (35\%) & 4\% & 51\% & 97\% & 98\% & 183/190 \\
20\_rosids & 1.07\% & 2201 (45\%) & 8\% & 73\% & 90\% & 91\% & 184/190 \\
\end{longtable}
}

\textbf{Table D8. The distance-indexed solver} (Section 8.6). 10 threads. Configuration of Section 7.1, time limit 3600 s. Time in seconds. Last column: final gap of the runs that did not certify.

{\def\LTcaptype{none} 
\begin{longtable}[]{@{}llllll@{}}
\toprule\noalign{}
instance & \(n\) & certified & time (s) & nodes & final gap \\
\midrule\noalign{}
\endhead
\bottomrule\noalign{}
\endlastfoot
Primates12 & 12 & 3 of 3 & 0.6 & 1 & -- \\
M17 & 17 & 3 of 3 & 7.7 {[}6.8--8.9{]} & 1 & -- \\
M18 & 18 & 3 of 3 & 94.5 {[}78.4--103{]} & 2081 {[}1717--2164{]} & -- \\
20\_euros2 & 20 & 3 of 3 & 6.5 {[}6.4--6.6{]} & 1 & -- \\
21\_nucleic & 21 & 3 of 3 & 253 {[}189--269{]} & 597 {[}391--780{]} & -- \\
22\_euros2 & 22 & 3 of 3 & 20.7 {[}20.5--25.5{]} & 1 & -- \\
23\_euros2 & 23 & 3 of 3 & 48.5 {[}45.1--53.4{]} & 1 & -- \\
24\_proteic & 24 & 3 of 3 & 1451 {[}620--2623{]} & 3479 {[}1286--6122{]} & -- \\
25\_proteic & 25 & 3 of 3 & 321 {[}285--435{]} & 852 {[}733--1314{]} & -- \\
26\_proteic & 26 & 3 of 3 & 763 {[}598--1501{]} & 1231 {[}881--3040{]} & -- \\
27\_nucleic & 27 & 3 of 3 & 886 {[}488--1543{]} & 1832 {[}870--2455{]} & -- \\
28\_nucleic & 28 & 2 of 3 & 3578 {[}3289--3600{]} & 3106 {[}2450--3301{]} & 0.049\% \\
29\_B-NS1 & 29 & 3 of 3 & 475 {[}362--604{]} & 319 {[}132--477{]} & -- \\
30\_B-NS1 & 30 & 3 of 3 & 918 {[}654--1099{]} & 449 {[}267--605{]} & -- \\
M43 & 43 & 0 of 1 & 3601 & 1 & 0.867\% \\
\end{longtable}
}

\textbf{Table D9. The distance-indexed model by cut configuration} (Section 8.7). 10 threads, except the two root-LP rows (one thread). (a) base, no separated cuts; (b) strong triangles strengthened by parity on all triples; then cumulatively generalized Kraft (GK), 2-K-split and circular order (2K, CO), and four-point (4pt); full is the configuration of Section 7.1, and full − CO the same without circular order. (a) and (b) are root LPs; the other rows are full solves (time limit 3600 s, three seeds where the first certifies), of which we report the converged root bound. On M18, the + GK run with seed 2 ended early, in each of three attempts, with an internal error of the solver after it rejected a heuristic solution; we report seeds 1 and 3 for that cell. Top: root gap to \(L^*\) (\%). Bottom: nodes to certification; ``non-tree'' marks runs that end at a matrix below \(L^*\) that is not a tree.

{\def\LTcaptype{none} 
\begin{longtable}[]{@{}lllll@{}}
\toprule\noalign{}
configuration & M17 & M18 & 20\_euros2 & 20\_rosids \\
\midrule\noalign{}
\endhead
\bottomrule\noalign{}
\endlastfoot
(a) base & 1.160 & 2.633 & 0.862 & 3.551 \\
(b) parity triangles & 1.160 & 2.633 & 0.862 & 3.551 \\
+ GK & 0.105 {[}0.102--0.106{]} & 0.756 {[}0.749--0.763{]} & 0.015 {[}0.015--0.017{]} & 0.394 {[}0.354--0.398{]} \\
+ 2K, CO & 0.105 {[}0.102--0.109{]} & 0.768 {[}0.758--0.771{]} & 0.012 {[}0.009--0.018{]} & 0.391 \\
+ 4pt & 0.009 {[}0.005--0.009{]} & 0.607 {[}0.603--0.621{]} & 0.019 {[}0.016--0.024{]} & 0.073 {[}0.070--0.073{]} \\
full & 0.010 {[}0.010--0.011{]} & 0.653 {[}0.631--0.661{]} & 0.008 {[}0.003--0.011{]} & 0.076 {[}0.071--0.086{]} \\
full − CO & 0.010 {[}0.010--0.011{]} & 0.653 {[}0.631--0.661{]} & 0.008 {[}0.003--0.011{]} & 0.076 {[}0.071--0.086{]} \\
\end{longtable}
}

{\def\LTcaptype{none} 
\begin{longtable}[]{@{}lllll@{}}
\toprule\noalign{}
configuration & M17 & M18 & 20\_euros2 & 20\_rosids \\
\midrule\noalign{}
\endhead
\bottomrule\noalign{}
\endlastfoot
(a) base & -- & -- & -- & -- \\
(b) parity triangles & -- & -- & -- & -- \\
+ GK & 104 {[}97--119{]} & non-tree & 1 & non-tree \\
+ 2K, CO & 71 {[}36--130{]} & 5263 {[}3318--5472{]} & 1 & 0 of 1 certified \\
+ 4pt & 1 & 965 {[}619--1306{]} & 1 & 131 {[}115--195{]} \\
full & 1 & 2081 {[}1717--2164{]} & 1 & 230 {[}127--376{]} \\
full − CO & 1 & 1700 {[}1321--3024{]} & 1 & 240 {[}139--409{]} \\
\end{longtable}
}

\textbf{Table D10. Rows of the distance-indexed base model added to the compact root relaxation} (Section 8.7). One thread. With the compact model's manifold imposed exactly (Section 7.2). Root gap to \(L^*\) (\%); a cell marked * is the bound of the arm's last solve before the conic solver failed, which is valid but may be weaker. L: the level indicators; M: the manifold equality; T: the triangle inequalities on all triples; 2K: lifted 2-K-splits for every \(K\); 2K\(_4\): the distance-indexed solver's root 2-K-split separation, \(K\le4\); 4pt\(_1\): the static four-point rows of every quartet containing leaf 1. Relaxations with M are linear programs (M implies the compact model's convex manifold inequality) and are solved by dual simplex with the safe bound; the others by MOSEK with the rigorous dual bound (Section 7.2).

{\def\LTcaptype{none} 
\begin{longtable}[]{@{}llllll@{}}
\toprule\noalign{}
relaxation & Primates12 & M17 & M18 & 20\_euros2 & 20\_rosids \\
\midrule\noalign{}
\endhead
\bottomrule\noalign{}
\endlastfoot
compact & 0.642 & 0.805 & 2.607 & 0.558 & 1.066 \\
+ L & 0.642 & 0.805 & 2.607 & 0.558 & 1.066 \\
+ L + M & 0.343 & 0.616 & 2.331 & 0.397 & 0.999 \\
+ L + T & 0.642 & 0.805 & 2.607 & 0.558 & 1.066 \\
+ L + M + T & 0.065 & 0.424 & 1.867 & 0.292 & 0.817 \\
+ L + M + T + 2K & 0.032 & 0.413 & 1.856 & 0.285 & 0.815 \\
+ L + 2K\(_4\) & 0.642 & 0.805 & 2.607 & 0.558 & 1.066 \\
+ L + M + T + 4pt\(_1\) & 0.065 & 0.424 & 1.867 & 0.292 & 0.817 \\
\end{longtable}
}

\textbf{Table D11. Where the distance-indexed model's root bound comes from} (Section 8.7). One thread. Root gap to \(L^*\) of the distance-indexed model in the configuration with generalized Kraft only (Table D9), stopped at the root node, with Gurobi's general-purpose cutting planes off (Cuts 0) and on (default), one thread; and, for comparison, the compact relaxation with the distance-indexed model's level indicators, manifold equality, triangles and lifted 2-K-splits added (Table D10), and the compact relaxation alone, both with the manifold imposed exactly.

{\def\LTcaptype{none} 
\begin{longtable}[]{@{}lllll@{}}
\toprule\noalign{}
instance & Gurobi cuts off & Gurobi cuts on & compact + base rows & compact \\
\midrule\noalign{}
\endhead
\bottomrule\noalign{}
\endlastfoot
M17 & 0.486\% & 0.105\% & 0.413\% & 0.805\% \\
M18 & 2.579\% & 0.763\% & 1.856\% & 2.607\% \\
20\_euros2 & 0.345\% & 0.017\% & 0.285\% & 0.558\% \\
20\_rosids & 0.987\% & 0.398\% & 0.815\% & 1.066\% \\
\end{longtable}
}

\textbf{Table D12. The distance-indexed model's static root in two MIP solvers} (Section 8.7). One thread. The model of the generalized-Kraft configuration exported before any callback cut (so without generalized Kraft), solved on one thread to the end of the root node with each solver's general-purpose cutting planes at their defaults and at their most aggressive settings (Gurobi Cuts 3; every CPLEX cut family at its upper limit). Root gap to \(L^*\) (\%). Last column: CPLEX with every cut family off; Gurobi with Cuts 0 stays within 0.12 points of the LP value.

{\def\LTcaptype{none} 
\begin{longtable}[]{@{}lllllll@{}}
\toprule\noalign{}
instance & LP & Gurobi & Gurobi aggr. & CPLEX & CPLEX aggr. & CPLEX off \\
\midrule\noalign{}
\endhead
\bottomrule\noalign{}
\endlastfoot
M17 & 1.160 & 0.548 & 0.539 & 0.397 & 0.357 & 1.160 \\
M18 & 2.633 & 0.897 & 0.801 & 1.066 & 0.918 & 2.633 \\
20\_euros2 & 0.862 & 0.381 & 0.355 & 0.342 & 0.302 & 0.862 \\
20\_rosids & 3.551 & 3.114 & 2.073 & 0.519 & 0.562 & 3.551 \\
28\_nucleic & 2.838 & 2.213 & 2.164 & 2.163 & 2.102 & 2.838 \\
\end{longtable}
}

\section{Appendix E. Numerical reliability}\label{appendix-e.-numerical-reliability}

The tolerances and parameters of Sections 7.1 and 7.2 take the following values.

\begin{itemize}
\tightlist
\item
  \textbf{Compact solver:} separation tolerance \(\epsilon_{\mathrm{sep}}=10^{-7}\); absolute pruning tolerance \(\epsilon_{\mathrm{prune}}=10^{-6}\), on objective values between about \(0.08\) and \(7\); CPLEX's feasibility and optimality tolerances \(\epsilon_{\mathrm{LP}}=10^{-9}\); the largest number of clusters whose trees are enumerated exactly, \(m_{\mathrm{enum}}=10\).
\item
  \textbf{Distance-indexed solver:} relative MIP gap \(10^{-8}\), set by the implementation; Gurobi's defaults otherwise: absolute feasibility and optimality tolerances of \(10^{-6}\), an absolute integrality tolerance of \(10^{-5}\) and an absolute MIP gap of \(10^{-10}\).
\end{itemize}

Both solvers compute in floating point, and the distance-indexed model's objective coefficients \(2^{-\ell}\) span a factor \(2^{n-3}\), which exceeds \(10^8\) for \(n\ge30\). Catanzaro, Frohn and Pesenti {[}35{]} show that the objective itself becomes numerically unstable once its smallest terms \(D_{ij}2^{-\tau_{ij}}\) approach machine precision, and that no rescaling of the exponents \(\tau\) avoids this without changing the optimal tree. On every instance that either solver certifies here, the smallest term is at least \(2\times10^{-11}\) of the optimal value; on M43 it is \(9\times10^{-16}\), a few units of machine precision. (The distance-indexed solver multiplies the objective by \(2^{n+1}\); scaling by a constant changes neither the optimum nor any condition number.) This appendix tests whether rounding could have changed any reported result, in the objective or in the relaxations.

\textbf{Every certified optimum is exactly a tree's value.} For each instance of Table D8, every certified run of the distance-indexed solver (Tables D8 and D9, configurations that separate the four-point condition) returns the same tree. Its path-length matrix satisfies the four-point condition, and its BME value, computed in exact rational arithmetic from the decimal input, differs from the value the solver reports by at most a relative \(3\times10^{-12}\). Its value is therefore exact. That no better tree exists rests, as for most computational results in exact combinatorial optimization, on three things: the validity of the model's rows and cut families for every tree, proved by Catanzaro \emph{et al.} {[}12{]}, {[}13{]}; their correct implementation; and the correctness of the search up to the solver's tolerances. Independent certification of the bounds and of each cut has been carried out for the travelling salesman problem {[}38{]}, and exact MIP solving with verifiable
certificates {[}39{]} covers the search; a formal proof of a BME optimum at these sizes would need both. The checks below test the three ingredients where they can, without amounting to such a proof: the enumeration covers only \(n\le12\), and the exact row checks show that no row cuts off the certified tree, not that none cuts off a better one.

\textbf{Safe bounds and rows that would cut off the optimum.} Every LP bound in Appendix D comes with a safe lower bound in the sense of Neumaier and Shcherbina {[}34{]}. For the LP \(\min\{c^\top x: Ax\ \square\ b,\ l\le x\le u\}\) and any dual vector \(y\) of the correct signs, \(b^\top y+\sum_j\min(r_jl_j,r_ju_j)\) with \(r=c-A^\top y\) is a lower bound on the LP optimum. We take \(y\) from the solver, set any component of the wrong sign to zero, and evaluate the bound in exact rational arithmetic on the LP as written in double precision, so the result is rigorous for that LP. Separately, we evaluated every row of each final LP, again exactly, at the point of the certified optimal tree: a row violated there would make the bound invalid for BME, whatever the arithmetic. Table E1 shows that no row of any of these LPs cuts off the optimal tree, and that the safe bounds lie below the solver's values by at most a relative \(10^{-14}\), far less than any gap we report.

\textbf{Table E1. Safe bounds and the exact check at the optimal tree.} One thread. For each group of LP bounds: the number of LPs, the most negative relative difference between the safe bound and the solver's LP value (negative: the floating-point value overstates the bound), the number of LP rows that cut off the certified optimal tree in exact arithmetic, and the number of LPs whose safe bound is \(-\infty\).

{\def\LTcaptype{none} 
\begin{longtable}[]{@{}lllll@{}}
\toprule\noalign{}
result & LPs & safe vs LP & rows cutting off \(T^*\) & infinite \\
\midrule\noalign{}
\endhead
\bottomrule\noalign{}
\endlastfoot
Table D3 & 15 & -5e-16 & 0 & 0 \\
Table D5 & 16 & -9e-16 & 0 & 0 \\
Table D6 & 8 & -1e-14 & 0 & 0 \\
Section 8.4, odd cycles & 16 & -1e-14 & 0 & 0 \\
Table D10 & 20 & -- & 0 & 0 \\
\end{longtable}
}

\textbf{Conic bounds in double precision.} The branch and bound evaluates the Lagrangian bound of Section 7.2 at every node in double precision, with explicit bounds on the rounding errors in the style of Higham {[}40{]}. Let \(\delta=2^{-53}\) be the unit roundoff and \(\gamma_k=k\delta/(1-k\delta)\); a sum or dot product of \(k\) terms, computed in any order, is within \(\gamma_k\) times the sum of the absolute values of its terms. With \(m\) rows, the computed reduced cost \(\hat r_j\) is therefore within \(\varepsilon_j=\gamma_{m+2}(|c_j|+(|A|^\top|y|)_j)\) of \(r_j\). The solver doubles \(\varepsilon_j\), to cover the rounding in computing it, and subtracts it from \(\hat r_j\), which gives a value \(\tilde r_j\le r_j\). Each one-dimensional minimum \(\varphi_j(r)=\min_{l_j\le x\le u_j}(rx+\mu x\ln x)\) is nondecreasing in \(r\), since its derivative is the minimiser \(x\ge l_j>0\), so \(\varphi_j(\tilde r_j)\le\varphi_j(r_j)\). For \(\mu>0\) the stationary point is \(e^{t}\) with
\(t=-\tilde r_j/\mu-1\). The solver takes the minimum at the endpoint \(l_j\) only when \(t<\ln l_j\) by more than the rounding error in \(t\) plus \(10^{-6}\), at \(u_j\) only when \(t>\ln u_j\) by as much, and otherwise takes the minimum over all \(x>0\), which is \(-\mu e^{t}\) and never above the minimum over the box. Each term is then lowered by 16 units of roundoff of the magnitudes it is computed from, which covers exponentials and logarithms accurate to a few units in the last place, and each sum by \(\gamma\) of its length. The result is at most the exact value of the Lagrangian bound for the multipliers in hand, given IEEE 754 arithmetic and that accuracy of the exponential and logarithm.

\textbf{Certificates against exhaustive enumeration.} On instances small enough to enumerate every tree, a certificate can be compared with the truth. We cropped each real instance to \(n=6,\dots,12\) leaves, both to its first \(n\) leaves and to three random subsets of \(n\) leaves per size, enumerated all \((2n-5)!!\) trees by stepwise addition {[}41{]}, re-ranked the best 200 in exact arithmetic, and ran both solvers on each crop with one thread. These crops test correctness, not scaling. Over the 196 crops of the seven real instances (at each \(n=6,\dots,12\), the first \(n\) leaves and three random subsets) and the five synthetic instances, both solvers certified all 201. No certificate from either solver is false: every certified tree is optimal, including on crops where the best and second-best trees differ by a relative \(10^{-5}\) (Table E2).

\textbf{Table E2. Certificates against exhaustive enumeration.} One thread per crop; not comparable with the times of Tables D1 and D8. Crops of each instance to \(n=6,\dots,12\) leaves (the first \(n\), or three random subsets per \(n\)). Columns: crops, crops certified by each solver, smallest relative gap between the optimum and the best non-optimal tree, and false certificates (a certified tree that is not optimal). Both solvers had a time limit of one hour per crop.

{\def\LTcaptype{none} 
\begin{longtable}[]{@{}llllll@{}}
\toprule\noalign{}
instance & crops & distance-indexed & compact & runner-up gap & false certificates \\
\midrule\noalign{}
\endhead
\bottomrule\noalign{}
\endlastfoot
20\_euros2 & 7 & 7 & 7 & 2e-05 & 0 \\
20\_euros2 (random) & 21 & 21 & 21 & 2e-05 & 0 \\
20\_rosids & 7 & 7 & 7 & 4e-05 & 0 \\
20\_rosids (random) & 21 & 21 & 21 & 3e-05 & 0 \\
30\_B-NS1 & 7 & 7 & 7 & 5e-03 & 0 \\
30\_B-NS1 (random) & 21 & 21 & 21 & 1e-04 & 0 \\
M17 & 7 & 7 & 7 & 8e-04 & 0 \\
M17 (random) & 21 & 21 & 21 & 1e-04 & 0 \\
M18 & 7 & 7 & 7 & 2e-04 & 0 \\
M18 (random) & 21 & 21 & 21 & 1e-04 & 0 \\
M43 & 7 & 7 & 7 & 5e-04 & 0 \\
M43 (random) & 21 & 21 & 21 & 1e-05 & 0 \\
Primates12 & 7 & 7 & 7 & 1e-03 & 0 \\
Primates12 (random) & 21 & 21 & 21 & 2e-03 & 0 \\
synthetic & 5 & 5 & 5 & 1e-03 & 0 \\
\end{longtable}
}

\textbf{Conditioning.} The condition number of a basis, the ratio of its largest to its smallest singular value, bounds how much rounding errors in the data can be amplified in the solution of the corresponding linear system {[}40{]}. It is unchanged when the whole matrix is multiplied by a scalar but not when rows and columns are scaled separately, so we report it for the optimal root basis after Ruiz's equilibration {[}42{]}, which divides every row and column by the square root of its largest absolute entry, repeatedly, until all of them are within \(10^{-6}\) of \(1\). Gurobi also reports a condition number for the optimal basis (its \texttt{KappaExact} attribute); its documentation does not say whether this refers to the model as written or to Gurobi's internally scaled model, so we computed our own from the basis that Gurobi reports. We measured the root LP of the compact model without the manifold constraint, and the static model of the distance-indexed solver before any
callback cut, on the first \(n\) leaves of 30\_B-NS1 and M43 (Table E3). The compact model's root basis stays below about \(10^5\) up to \(n=43\), and the distance-indexed static model grows steadily, to about \(4\times10^6\) at \(n=43\).

\textbf{Table E3. Condition number of the optimal root basis}. One thread. First \(n\) leaves: Gurobi's \texttt{KappaExact}, and in parentheses our own 2-norm value after Ruiz equilibration where the basis is small enough for a dense decomposition. ``Compact'' is the root LP of the compact model without the manifold constraint; ``distance-indexed'' the static model of Catanzaro \emph{et al.}

\emph{30\_B-NS1}

{\def\LTcaptype{none} 
\begin{longtable}[]{@{}lll@{}}
\toprule\noalign{}
\(n\) & compact & distance-indexed \\
\midrule\noalign{}
\endhead
\bottomrule\noalign{}
\endlastfoot
6 & 3e+01 (4e+01) & 7e+02 (2e+03) \\
8 & 4e+02 (2e+02) & 2e+03 (4e+03) \\
10 & 1e+03 (6e+02) & 3e+03 (5e+03) \\
12 & 4e+03 (2e+03) & 2e+04 (9e+03) \\
15 & 7e+04 (2e+04) & 2e+04 (3e+04) \\
20 & 5e+04 (2e+04) & 1e+05 \\
25 & 6e+03 & 5e+05 \\
30 & 4e+03 & 5e+05 \\
\end{longtable}
}

\emph{M43}

{\def\LTcaptype{none} 
\begin{longtable}[]{@{}lll@{}}
\toprule\noalign{}
\(n\) & compact & distance-indexed \\
\midrule\noalign{}
\endhead
\bottomrule\noalign{}
\endlastfoot
6 & 5e+01 (5e+01) & 3e+03 (4e+03) \\
8 & 5e+02 (3e+02) & 8e+03 (6e+03) \\
10 & 1e+03 (6e+02) & 4e+04 (2e+04) \\
12 & 1e+03 (6e+02) & 2e+04 (4e+04) \\
15 & 8e+02 (6e+02) & 7e+04 (4e+04) \\
20 & 5e+03 (3e+03) & 2e+05 \\
25 & 3e+03 & 3e+05 \\
30 & 2e+03 & 9e+05 \\
35 & 3e+03 & 2e+06 \\
40 & 7e+03 & 1e+06 \\
43 & 9e+03 & 4e+06 \\
\end{longtable}
}

\textbf{Robustness of the distance-indexed solver's certificates.} On 29\_B-NS1 and 30\_B-NS1, the largest instances that certify, we reran the distance-indexed solver with each separated cut family except four-point switched off in turn, and with Gurobi's numerical emphasis at its maximum (\texttt{NumericFocus} 3) and its absolute feasibility, optimality and integrality tolerances lowered to \(10^{-9}\). A cut that wrongly removed the optimal tree through rounding would show as a different optimum when its family is off, and a certificate that depended on the tolerances would show as a different optimum when they are tightened. Every run that certified, six of the ten, returned the same tree as the full configuration, and the four that did not certify within the hour held that tree as their incumbent (Table E4). Switching off generalized Kraft leaves gaps of \(2.7\%\) and \(4.7\%\); switching off the weak four-point cuts or tightening the tolerances slows certification without
changing its result. No cut family, and no tolerance in this range, changes the certified optimum.

\textbf{Table E4. The distance-indexed solver with one cut family switched off, or with strict tolerances.} 10 threads. Seed 1, time limit 3600 s. Strict tolerances: Gurobi's numerical emphasis at its maximum (NumericFocus 3) and its feasibility, optimality and integrality tolerances at \(10^{-9}\). Last column: the best tree found, compared with the tree of the full configuration (Table D8).

{\def\LTcaptype{none} 
\begin{longtable}[]{@{}lllll@{}}
\toprule\noalign{}
configuration & instance & certified & time (s) & tree \\
\midrule\noalign{}
\endhead
\bottomrule\noalign{}
\endlastfoot
without generalized Kraft & 29\_B-NS1 & no (2.73\%) & 3600 & same \\
without generalized Kraft & 30\_B-NS1 & no (4.66\%) & 3600 & same \\
without 2-K-splits & 29\_B-NS1 & yes & 383 & same \\
without 2-K-splits & 30\_B-NS1 & yes & 1012 & same \\
without circular order & 29\_B-NS1 & yes & 594 & same \\
without circular order & 30\_B-NS1 & yes & 1086 & same \\
without weak four-point & 29\_B-NS1 & yes & 1470 & same \\
without weak four-point & 30\_B-NS1 & no (0.14\%) & 3601 & same \\
strict tolerances & 29\_B-NS1 & yes & 2016 & same \\
strict tolerances & 30\_B-NS1 & no (0.15\%) & 3601 & same \\
\end{longtable}
}

\section{Appendix F. The branch and bound}\label{appendix-f.-the-branch-and-bound}

This appendix states the compact solver's branch and bound (Section 7.2) in full. A node is a pair \((C,F)\) of \textbf{committed} splits \(C\), imposed as \(W[S]=\tfrac12\), and \textbf{forbidden} splits \(F\), imposed as \(W[S]\ge\tfrac34\); its \textbf{face} is the set of binary trees whose splits include \(C\) and exclude \(F\). Two invariants hold at every node: \(C\) is laminar, because the search never commits a split that crosses a committed one, and \(C\cap
F=\emptyset\), because the split it branches on is always outside \(C\cup F\). Comparisons of objective values use the tolerance of Section 7.1.

\subsection{F.1 Repeated cherry contraction}\label{f.1-repeated-cherry-contraction}

Contract merges two clusters of leaves into one whenever a test says they form a cherry, and repeats on the merged clusters until no pair passes or three clusters remain. It serves three purposes, with two tests. The \textbf{committed test} merges \(X\) and \(Y\) when \(X\cup Y\) is a side of a split in \(C\); it contracts exactly what every tree of the face shares. The \textbf{cherry test} merges \(X\) and \(Y\) when \(\hat w_{XY}=\tfrac14\) and \(\hat w_{XK}=\hat
w_{YK}\) for every other cluster \(K\), where \(\hat w\) is \(w\) updated at each merge by \(\hat w_{ZK}=2\hat w_{XK}\): if \(X\) and \(Y\) form a cherry merged into \(Z\), the path from a leaf of \(X\) to \(K\) is one edge longer than the path from \(Z\) to \(K\). The cherry test recognises a tree: if it stops at three clusters, its recorded splits are the \(n-3\) splits of a tree \(T\), which the search accepts only if \(w_T=w\), since the test checks cherry pairs only.

\begin{algorithm}[H]
\small
\caption{\textsc{Contract}$(\mathit{test})$}\label{alg:contract}
\begin{algorithmic}[1]
\State clusters $\gets\{\{1\},\dots,\{n\}\}$; $e_i\gets0$ for every leaf $i$; $\hat w\gets w$
\Repeat
  \If{two clusters $X,Y$ pass $\mathit{test}(X,Y)$}
    \State replace $X$ and $Y$ by $Z=X\cup Y$; $e_i\gets e_i+1$ for every $i\in Z$
    \State $\hat w_{ZK}\gets2\hat w_{XK}$ for every other cluster $K$
    \State record $Z$ as a split
  \EndIf
\Until{no pair passes, or $3$ clusters remain}
\State \Return clusters, $e$, the recorded splits
\Statex
\Statex \emph{committed test:} $X\cup Y$ is a side of a split in $C$
\Statex \emph{cherry test:} $\hat w_{XY}=\tfrac14$ and $\hat w_{XK}=\hat w_{YK}$ for every other cluster $K$
\end{algorithmic}
\end{algorithm}

With the committed test, the objective of every tree in the face splits into a part that is the same for all of them and a smaller instance on the clusters.

\textbf{Lemma F.1.} Let \(C\) be laminar, and let Contract with the committed test return the clusters \(A_1,\dots,A_m\) (\(m\ge3\)) and the values \(e_i\). Let \(C'\) be the splits of \(C\) it did not record, and \(F'\) the splits of \(F\) whose sides are unions of clusters. Then:

\begin{enumerate}
\def\labelenumi{\arabic{enumi}.}
\tightlist
\item
  every tree \(T\) whose splits include \(C\) contains each cluster \(A\) as the leaf set of a subtree hanging from one edge, with a shape that does not depend on \(T\), and \(e_i\) is the number of edges from leaf \(i\) to the vertex \(r_A\) of that subtree nearest the rest of the tree;
\item
  replacing each such subtree by one leaf gives a binary tree \(T'\) on the clusters, and \(T\mapsto T'\) is a bijection from the face of \((C,F)\) onto the trees on the clusters whose splits include \(C'\) and exclude \(F'\), provided \(C\cap F=\emptyset\);
\item
  for \(i\in A\) and \(j\in B\ne A\), \(\tau_{ij}=e_i+e_j+\tau'_{AB}\), where \(\tau'\) are the path lengths of \(T'\); hence \[L(T)=K+\sum_{A<B}2^{-\tau'_{AB}}D'_{AB},\qquad
  D'_{AB}=\sum_{i\in A,\,j\in B}2^{-e_i-e_j}D_{ij},\] where \(K=\sum_A\sum_{i<j\in A}2^{-\tau_{ij}}D_{ij}\) is the same for every tree of the face.
\end{enumerate}

\emph{Proof.} (1) By induction on the merges. Initially every cluster is a leaf \(i\), with \(r_{\{i\}}=i\) and \(e_i=0\). Suppose the claim holds for the clusters \(X\) and \(Y\) that are merged, and let \(Z=X\cup Y\), a side of a split in \(C\), hence of a split of \(T\); \(Z\) is the leaf set of a subtree of \(T\) hanging from one edge. Its vertex \(r_Z\) nearest the rest of the tree has two child subtrees, whose leaf sets \(Z_1,Z_2\) partition \(Z\). The leaf set \(X\) of a subtree inside \(Z\) lies within \(Z_1\) or within \(Z_2\), and so does \(Y\); as \(X\cup Y=Z\), one of them is \(Z_1\) and the other \(Z_2\). So \(r_Z\) is adjacent to \(r_X\) and \(r_Y\), the shape of the subtree on \(Z\) is fixed, and each leaf of \(Z\) is one edge further from \(r_Z\) than from its old vertex, which is the update \(e_i\leftarrow e_i+1\).

(2) The splits of \(T\) are those of \(T'\), all unions of clusters, together with the splits that separate a subtree inside a cluster, which are exactly the recorded ones, since each merge creates one and a binary subtree on \(k\) leaves has \(k-1\) such subtrees of size at least \(2\). Gluing the fixed subtrees back onto the leaves of \(T'\) inverts the map. A split of \(C\) is either recorded, and present in every \(T\), or not; in the latter case it is compatible with every recorded split, so each side is a union of clusters or lies inside one, and inside a cluster it would cross a recorded split or be one, as the recorded splits inside a cluster are all its subtree splits. So \(T\) contains \(C\) if and only if \(T'\) contains \(C'\). A split of \(F\) that is not a union of clusters separates a cluster: it is either a recorded split, excluded by \(C\cap F=\emptyset\), or crosses one, and then no tree of the face contains it. So \(T\) excludes \(F\) if and only if \(T'\) excludes
\(F'\).

(3) The path from \(i\in A\) to \(j\in B\) runs from \(i\) to \(r_A\) inside the subtree on \(A\) (\(e_i\) edges), from \(r_A\) to \(r_B\) along the path between the leaves \(A\) and \(B\) of \(T'\) (\(\tau'_{AB}\) edges), and from \(r_B\) to \(j\) (\(e_j\) edges). Substituting into \(L(T)=\sum_{i<j}2^{-\tau_{ij}}D_{ij}\) and grouping the pairs by cluster gives the formula; pairs inside one cluster have path lengths fixed by (1). \(\square\)

The search enumerates a face exactly when at most \(m_{\mathrm{enum}}=10\) clusters remain: at most \((2m_{\mathrm{enum}}-5)!!=2{,}027{,}025\) trees on the clusters, generated by stepwise addition (leaf \(k\) inserted on each edge of every tree on the first \(k-1\)), in compiled code.

\subsection{F.2 The algorithms}\label{f.2-the-algorithms}

\begin{algorithm}[H]
\small
\caption{\textsc{BranchAndBound}$(D)$}\label{alg:bnb}
\begin{algorithmic}[1]
\State $T^*\gets$ FastME tree on $D$, improved by NNI; $U\gets L(T^*)$
\State $(\beta,w)\gets\Call{NodeBound}{\emptyset,\emptyset}$; $Q\gets\{(\beta,\emptyset,\emptyset,w)\}$\Comment{priority queue on $\beta$}
\State $\mathit{next}\gets\sv{none}$\Comment{an inherited child, taken first}
\While{($\mathit{next}\ne\sv{none}$ or $Q\ne\emptyset$) and the budget is not exhausted}
  \If{$\mathit{next}\ne\sv{none}$}
    \State $(\beta,C,F,w)\gets\mathit{next}$; $\mathit{next}\gets\sv{none}$
    \IIf{$\beta\ge U$}{\Continue}
  \Else
    \State $(\beta,C,F,w)\gets$ remove the entry of least $\beta$ from $Q$
    \IIf{$\beta\ge U$}{\Return $T^*$, certified}\Comment{every open node is at least $U$}
  \EndIf
  \If{$\Call{Contract}{\text{cherry test}}$ recognises a tree $T$ with $w_T=w$ and $L(T)=\beta$}
    \IIf{$\beta<U$}{$T^*\gets T$; $U\gets\beta$}
    \State \Continue\Comment{closed: the LP optimum is a tree}
  \EndIf
  \State $S\gets$ the most balanced split of $T^*$ with $\tfrac12<W[S]<\tfrac34$, not in $C\cup F$
  \IIf{there is none}{$S\gets$ a split with $\tfrac12<W[S]<\tfrac34$, not in $C\cup F$}
  \If{there is none}
    \State $(\mathit{status},X)\gets\Call{SettleFace}{w,C,F}$
    \If{$\mathit{status}=\sv{tree}$}
      \IIf{$L(X)<U$}{$T^*\gets X$; $U\gets L(X)$}
      \State \Continue\Comment{closed exactly}
    \EndIf
    \IIf{$\mathit{status}=\sv{empty}$}{\Continue}\Comment{closed: no tree}
    \State $\Call{IncumbentSearch}{w,C,F}$
    \IIf{$\mathit{status}=\sv{branch}$}{$S\gets X$}
  \EndIf
  \If{SettleFace returned \sv{cherries}}\Comment{$X=(P_1,\dots,P_r)$}
    \State $\mathit{children}\gets\{(C\cup\{P_k\},\ F\cup\{P_1,\dots,P_{k-1}\}):k=1,\dots,r\}$
  \ElsIf{$S$ crosses a split of $C$}\Comment{no tree of the face contains $S$}
    \State $\mathit{children}\gets\{(C,\ F\cup\{S\})\}$
  \Else
    \State $\mathit{children}\gets\{(C,\ F\cup\{S\}),\ (C\cup\{S\},\ F)\}$\Comment{$W[S]\ge\tfrac34$, then $W[S]=\tfrac12$}
  \EndIf
  \For{$(C',F')\in\mathit{children}$}
    \If{$w$ satisfies the rows of $(C',F')$ and its cluster equalities}
      \State $(\beta',w')\gets(\beta,w)$\Comment{inherited, not solved}
      \IIf{$\mathit{next}=\sv{none}$}{$\mathit{next}\gets(\beta',C',F',w')$; \Continue}
    \Else
      \State $(\beta',w')\gets\Call{NodeBound}{C',F'}$
      \IIf{the relaxation is infeasible}{\Continue}
    \EndIf
    \IIf{$\beta'<U$}{add $(\beta',C',F',w')$ to $Q$}
  \EndFor
\EndWhile
\State \Return $T^*$, lower bound $\min\{\beta:(\beta,\cdot)\in Q\cup\{\mathit{next}\}\}$
\end{algorithmic}
\end{algorithm}

\begin{algorithm}[H]
\small
\caption{\textsc{SettleFace}$(w,C,F)$, with $C$ laminar and $C\cap F=\emptyset$}\label{alg:settle}
\begin{algorithmic}[1]
\State $(\text{clusters},e,\text{recorded})\gets\Call{Contract}{\text{committed test}}$; $m\gets|\text{clusters}|$
\If{$m\le m_{\mathrm{enum}}$}
  \State enumerate the trees on the clusters whose splits include $C'$ and exclude $F'$, and
  \Statex\hspace{\algorithmicindent}\hspace{\algorithmicindent} evaluate $L$ by Lemma F.1
  \IIf{there is none}{\Return \sv{empty}}
  \State \Return \sv{tree}, the best one
\EndIf
\State $S\gets$ the most balanced undecided split (not in $C\cup F$) with $W[S]=\tfrac12$ that
\Statex\hspace{\algorithmicindent} crosses no split of $C$\Comment{the half-cuts are listed as minimum cuts}
\IIf{there is one}{\Return \sv{branch}, $S$}
\State $V\gets$ the smallest set of clusters attached to one vertex of the skeleton (the tree on the
\Statex\hspace{\algorithmicindent} clusters whose edges are the splits of $C'$) that ends a single skeleton edge;
\Statex\hspace{\algorithmicindent} all the clusters if $C'=\emptyset$
\State $P_1,\dots,P_r\gets$ the pairs of clusters of $V$ whose union is not in $F$, ordered by the total
\Statex\hspace{\algorithmicindent} weight of $w$ between the two clusters, largest first
\IIf{$r=0$}{\Return \sv{empty}}\Comment{no pair of $V$ can form a cherry}
\State \Return \sv{cherries}, $(P_1,\dots,P_r)$\Comment{each $P_k$ as the split it commits}
\end{algorithmic}
\end{algorithm}

\begin{algorithm}[H]
\small
\caption{\textsc{IncumbentSearch}$(w,C,F)$: candidate trees only, never a bound}\label{alg:incumbent}
\begin{algorithmic}[1]
\State $P\gets$ the splits recorded by $\Call{Contract}{\text{cherry test}}$
\If{a split of $P$ crosses a split of $C$, or lies in $F$}
  \State $X\gets$ NNI hill climb from a tree whose splits include $C$
\Else
  \State $(\mathit{status},X)\gets\Call{SettleFace}{w,C\cup P,F}$, without choosing a split to branch on
  \IIf{more than $m_{\mathrm{enum}}$ clusters remain}{$X\gets$ NNI hill climb from a tree whose splits include $C$}
\EndIf
\IIf{$X$ is a tree and $L(X)<U$}{$T^*\gets X$; $U\gets L(X)$}
\end{algorithmic}
\end{algorithm}

\begin{algorithm}[H]
\small
\caption{\textsc{NodeBound}$(C,F)$}\label{alg:nodebound}
\begin{algorithmic}[1]
\State add $W[S]=\tfrac12$ for $S\in C$ and $W[S]\ge\tfrac34$ for $S\in F$ to the LP
\State $(\text{clusters},e,\cdot)\gets\Call{Contract}{\text{committed test}}$
\For{each cluster $Z$, a fixed leaf $z$ of $Z$, every other leaf $i$ of $Z$ and every leaf $c\notin Z$}
  \State add $2^{e_i}w_{ic}=2^{e_z}w_{zc}$ to the LP\Comment{cluster equalities}
\EndFor
\For{every two leaves $i,j$ of one cluster, at path length $p_{ij}$ within it}
  \State add $w_{ij}=2^{-p_{ij}}$ to the LP
\EndFor
\State run the cut loop (minimum cut, perfect matching, crossing half-cuts) to convergence:
\Statex\hspace{\algorithmicindent} LP bound $\beta$, point $w$
\Repeat
  \State solve the same rows and $\sum_{i<j}\mathrm{entr}(w_{ij})\ge(n-\tfrac32)\ln2$ by the conic solver
  \IIf{the solver fails or reports infeasibility}{\Break}\Comment{keep the LP bound and point}
  \State $\beta_c\gets$ its rigorous Lagrangian bound; $w_c\gets$ its optimum
  \If{a separator finds a cut violated at $w_c$}
    \State add it; rerun the cut loop; update $\beta$ and $w$
  \Else
    \IIf{$\beta_c>\beta$}{$(\beta,w)\gets(\beta_c,w_c)$}
    \State \Break
  \EndIf
\Until{false}
\State remove the node's own rows\Comment{the cuts stay: they hold for every tree}
\State \Return $(\beta,w)$
\end{algorithmic}
\end{algorithm}

The two calls of SettleFace differ only in what they treat as committed. In Algorithm 2 it receives the node's own splits, so its enumeration covers the whole face and may close the node. In Algorithm 4 it also receives the cherries of the LP point, which the trees of the node need not share; its best tree can then only replace the incumbent, and the node stays open. A child whose relaxation is cut short by the time limit stays open with its parent's bound, which enters the lower bound; Algorithm 2 omits this case. A relaxation that the conic solver reports infeasible is never pruned on that report: an interior-point method gives no exact certificate, so the node keeps its LP bound.

In Algorithm 2 the cherry branching (status cherries) gives children that partition the face: a tree of the face has some pair of \(V\) as a cherry, because the clusters of \(V\) hang from the rest of the tree as one rooted binary subtree (or form the whole tree), and child \(k\) holds exactly the trees whose first such pair in the order is \(P_k\). An inherited child keeps its parent's point, which is optimal for the child because the child's relaxation is contained in the parent's; its cluster equalities are the only rows it adds that the parent's point might violate, so the test checks them.

\subsection{F.3 The parallel search}\label{f.3-the-parallel-search}

\begin{algorithm}[H]
\small
\caption{\textsc{ParallelBranchAndBound}$(D,k)$}\label{alg:parallel}
\begin{algorithmic}[1]
\State as Algorithm 2 up to the root; start $k$ workers, each with its own LP (and conic) model
\While{some worker is busy, or the budget is not exhausted and a node is available}
  \State hand each idle worker $i$ a node: its own inherited child $\mathit{next}_i$ if $\beta(\mathit{next}_i)<U$,
  \Statex\hspace{\algorithmicindent}\hspace{\algorithmicindent} otherwise the entry of least $\beta$ from $Q$ if that $\beta<U$
  \State wait until some worker $i$ returns its node
  \State merge: $U$ and $T^*$ if it found a better tree; its children into $Q$, except that one
  \Statex\hspace{\algorithmicindent}\hspace{\algorithmicindent} inherited child becomes $\mathit{next}_i$; its new cuts into every other worker's
  \Statex\hspace{\algorithmicindent}\hspace{\algorithmicindent} queue of rows, added to that worker's model before its next node
\EndWhile
\State certified if no node is being processed, no $\mathit{next}_i<U$ and every entry of $Q$ is at least $U$
\State lower bound: the least $\beta$ over $Q$, the $\mathit{next}_i$ and the nodes being processed
\end{algorithmic}
\end{algorithm}

A worker processes a node as one pass of the loop of Algorithm 2, without the queue. The cuts are valid for every tree, so a worker may add them at any time. A child's bound is at least its parent's, so a node still being processed cannot produce a child below the lower bound.

\section{Appendix G. The four-point condition on the quartets that contain one leaf}\label{appendix-g.-the-four-point-condition-on-the-quartets-that-contain-one-leaf}

Catanzaro \emph{et al.} {[}13{]}, thm. 2 and prop. 8 prove the following by a case analysis of 81 combinations of resolutions. We give a short proof. Throughout, a quartet \(\{a,b,c,d\}\) satisfies the \textbf{strong four-point condition} if, of its three pair sums \(\tau_{ab}+\tau_{cd}\), \(\tau_{ac}+\tau_{bd}\) and \(\tau_{ad}+\tau_{bc}\), the two largest are equal and exceed the smallest by at least \(2\); a triple \(\{i,j,k\}\) satisfies the \textbf{strong triangle inequalities} if \(\tau_{ik}+\tau_{kj}-\tau_{ij}\ge2\) for each of the three choices of the middle leaf \(k\).

\begin{quote}
\textbf{Proposition G.1.} Let \(\tau\) be a real symmetric \(n\times n\) matrix, \(n\ge4\). (a) If \(\tau\) satisfies the strong four-point condition on every quartet that contains leaf \(1\), it satisfies it on every quartet. (b) If, in addition, \(\tau\) satisfies the strong triangle inequalities on every triple that contains leaf \(1\), it satisfies them on every triple.
\end{quote}

\emph{Proof.} For distinct leaves \(i,j\ne1\) let \((i|j)=\tfrac12(\tau_{1i}+\tau_{1j}-\tau_{ij})\), the \textbf{Gromov product} of \(i\) and \(j\) seen from leaf \(1\).

\emph{Translation.} For distinct \(i,j,k\ne1\), with \(\sigma=\tau_{1i}+\tau_{1j}+\tau_{1k}\), the three pair sums of the quartet \(\{1,i,j,k\}\) are \[\tau_{1i}+\tau_{jk}=\sigma-2(j|k),\qquad\tau_{1j}+\tau_{ik}=\sigma-2(i|k),\qquad
\tau_{1k}+\tau_{ij}=\sigma-2(i|j).\] So the strong four-point condition on \(\{1,i,j,k\}\) says exactly:

\begin{quote}
\begin{enumerate}
\def\labelenumi{(\Alph{enumi})}
\setcounter{enumi}{15}
\tightlist
\item
  of the three products \((i|j)\), \((i|k)\), \((j|k)\), the two smallest are equal and the largest exceeds them by at least \(1\).
\end{enumerate}
\end{quote}

In particular the largest of the three is unique. (Condition (P) is the three-point condition of an ultrametric for \(-(i|j)\), so part (a) says that the matrix \(\tau_{ij}-\tau_{1i}-\tau_{1j}=-2(i|j)\), the Farris transform of \(\tau\) with respect to leaf \(1\), is an ultrametric up to an additive constant.)

\emph{Part (a).} Let \(i,j,k,l\ne1\) be distinct. With \(\sigma'=\tau_{1i}+\tau_{1j}+\tau_{1k}+\tau_{1l}\), the three pair sums of \(\{i,j,k,l\}\) are \(\sigma'-2A\), \(\sigma'-2B\) and \(\sigma'-2C\), where \[A=(i|j)+(k|l),\qquad B=(i|k)+(j|l),\qquad C=(i|l)+(j|k).\] The strong four-point condition on \(\{i,j,k,l\}\) says that the two smallest of \(A,B,C\) are equal and the largest exceeds them by at least \(1\). Since the claim is symmetric in the four leaves, we may label them so that \((i|j)\) is the largest of the six products. By (P) on \(\{i,j,k\}\), \((i|k)=(j|k)=:a\le(i|j)-1\), and by (P) on \(\{i,j,l\}\), \((i|l)=(j|l)=:b\le(i|j)-1\). Hence \(B=C=a+b\). Let \(c=(k|l)\), and apply (P) to \(\{i,k,l\}\), whose products are \(a\), \(b\) and \(c\). If \(a<b\), the two smallest must be \(a\) and \(c\), so \(c=a\) and \(A-B=(i|j)-b\ge1\). If \(a>b\), likewise \(c=b\) and \(A-B=(i|j)-a\ge1\). If \(a=b\), then \(c\) is the largest, \(c\ge a+1\), and
\(A-B=\bigl((i|j)-a\bigr)+(c-a)\ge2\). In every case \(A\ge B+1=C+1\).

\emph{Part (b).} The strong triangle inequalities on \(\{1,i,j\}\) with middle leaf \(i\) and \(j\) give \(\tau_{ij}\ge2+|\tau_{1i}-\tau_{1j}|\), that is, \((i|j)\le\min(\tau_{1i},\tau_{1j})-1\). Let \(i,j,p\ne1\) be distinct and take \(j\) as the middle leaf. Expanding each entry by the definition of the products, \[\tau_{ij}+\tau_{jp}-\tau_{ip}=2\bigl(\tau_{1j}-(i|j)-(j|p)+(i|p)\bigr),\] so we need \(\tau_{1j}-(i|j)-(j|p)+(i|p)\ge1\). Apply (P) to \(\{i,j,p\}\). If \((i|p)\) is the largest, then \((i|j)=(j|p)=:a\), \((i|p)\ge a+1\) and \(\tau_{1j}\ge a+1\), so the left-hand side is at least \(2\). If \((i|j)\) is the largest, then \((j|p)=(i|p)\) and the left-hand side is \(\tau_{1j}-(i|j)\ge1\); the case where \((j|p)\) is the largest is symmetric. The other two choices of the middle leaf follow by renaming. \(\blacksquare\)

Neither part uses integrality, and part (a) does not use the triangle inequalities.

\section{Appendix H. Clique covers of the crossing graph}\label{appendix-h.-clique-covers-of-the-crossing-graph}

This appendix proves part (b) of Proposition 4, describes the subfamily of the cover of part (a) mentioned in Section 5.1, and computes the lower bound \(88\) at \(n=8\).

\textbf{Intersecting antichains.} For a clique \(K\) of \(G\), let \(\mathcal F_K\) be the family of the smaller blocks of its bipartitions (for a bipartition with two blocks of size \(n/2\), the block not containing leaf \(1\)). Two crossing bipartitions have all four quadrants nonempty, so any two members of \(\mathcal F_K\) intersect and neither contains the other, and every member has at most \(n/2\) elements.

\begin{quote}
\textbf{Lemma H.1} {[}26{]}, {[}27{]}. Let \(\mathcal F\) be a family of subsets of \([n]\), each with at most \(n/2\) elements, any two of which intersect and neither of which contains the other. Then \(\sum_{A\in\mathcal F}1/\binom{n-1}{|A|-1}\le1\).
\end{quote}

\emph{Proof.} We use the circle method of Katona {[}43{]}. Let us place \([n]\) on a circle in one of the \((n-1)!\) cyclic orders, and call a member of \(\mathcal F\) an arc if its elements are consecutive. Let \(B=(b_1,\dots,b_k)\) be a shortest arc. Every other arc \(A\) meets \(B\) and is neither contained in \(B\) nor contains it, so \(A\cap B\) is \(\{b_1,\dots,b_i\}\) or \(\{b_{i+1},\dots,b_k\}\) for some \(1\le i\le k-1\); it cannot contain both \(b_1\) and \(b_k\), since it would then contain every element outside \(B\) as well, giving \(|A|\ge n-k+2>n/2\). In the first case \(A\) ends at \(b_i\), in the second it starts at \(b_{i+1}\). Two arcs with the same end, or the same start, are nested, so for each \(i\) at most one arc ends at \(b_i\) and at most one starts at \(b_{i+1}\). Not both: such arcs lie in the \(\lfloor n/2\rfloor\) positions ending at \(b_i\) and the \(\lfloor n/2\rfloor\) positions starting at \(b_{i+1}\) respectively, which are disjoint. So there are at
most \(k\) arcs, each with at least \(k\) elements, and \(\sum_{A\text{ arc}}1/|A|\le1\). A set with \(a\) elements is an arc in exactly \(a!\,(n-a)!\) cyclic orders, so summing over all cyclic orders gives \(\sum_{A\in\mathcal F}a!\,(n-a)!/a\le(n-1)!\), where \(a=|A|\), which is the claim. \(\blacksquare\)

\emph{Proof of Proposition 4(b).} Let \(I\) be the set of integers \(i\) with \(\max(2,n/4)\le i<n/2\), so \(m=|I|\), and let \(c_n=1-e^{-n/16}-2^{-n/4}\). For \(i<j\) in \(I\) let \(E_{ij}\) be the set of crossing pairs whose smaller blocks have sizes \(i\) and \(j\); since \(i,j<n/2\), the smaller block of each bipartition is unique. An \(i\)-set \(A\) and a \(j\)-set \(B\) give a crossing pair if and only if \(A\cap B\ne\emptyset\) and \(A\not\subseteq B\), since \(B\not\subseteq A\) (as \(j>i\)) and \(A\cup B\ne[n]\) (as \(i+j<n\)) hold automatically. For a fixed \(A\), a uniformly random \(j\)-set misses \(A\) with probability at most \((1-i/n)^j\le e^{-ij/n}\le e^{-n/16}\) and contains \(A\) with probability \(\binom ji/\binom ni\le(j/n)^i\le2^{-i}\le2^{-n/4}\). Since \(\binom ni=\tfrac ni\binom{n-1}{i-1}\ge2\binom{n-1}{i-1}\), and likewise for \(j\), \[|E_{ij}|\ \ge\ c_n\binom ni\binom nj\ \ge\ 4c_n\binom{n-1}{i-1}\binom{n-1}{j-1}.\] For a clique \(K\), let \(a_i(K)\) be the
number of its bipartitions whose smaller block has size \(i\) and \(\beta_i(K)=a_i(K)/\binom{n-1}{i-1}\), so that \(\sum_i\beta_i(K)\le1\) by Lemma H.1. The clique contains at most \(a_i(K)\,a_j(K)\) pairs of \(E_{ij}\). For an edge clique cover \(\mathcal K\) of \(G\), covering \(E_{ij}\) requires \(\sum_{K\in\mathcal K}a_i(K)a_j(K)\ge|E_{ij}|\), that is, \(\sum_{K\in\mathcal K}\beta_i(K)\beta_j(K)\ge4c_n\). Summing over the \(\binom m2\) pairs \(i<j\) in \(I\), and using \(\sum_{i<j}\beta_i(K)\beta_j(K)\le\tfrac12\bigl(\sum_i\beta_i(K)\bigr)^2
\le\tfrac12\) for each \(K\), gives \(|\mathcal K|/2\ge4c_n\binom m2\). Since \(m\ge n/4-1\) and \(c_n\to1\), this is at least \((1-o(1))\,n^2/4\). \(\blacksquare\)

\textbf{A smaller cover.} The subfamily of the cover of Proposition 4(a) consists of the cliques \(Q^{2,2}_{x,y}\), which do not depend on \(y\) (each is the set of \(2\)-blocks containing \(x\)), and the cliques \(Q^{j,k}_{x,y}\) with \(2\le j<k\le n/2\), except \(Q^{3,k}_{x,s(x)}\), where \(s\) is a fixed permutation of the leaves without fixed points, an involution when \(n\) is even. When \(k=n/2\), \(Q^{j,k}_{x,y}\) and \(Q^{j,k}_{y,x}\) coincide, as a block of size \(n/2\) containing \(x\) but not \(y\) is the complement of one containing \(y\) but not \(x\); the subfamily counts each such clique once. It is an edge clique cover. Let \(S,S'\) cross, with smaller blocks \(B,B'\) of sizes \(j\le k\). If \(j=k\), we take \(x\in B\cap B'\) and \(y\notin B\cup B'\); both \(B\) and \(B'\) are \(x\)-blocks of size \(k\) without \(y\), so both bipartitions lie in \(Q^{2,k}_{x,y}\). If \(j<k\), we take \(x\in B\cap B'\) and \(y\in B\setminus
B'\), and both bipartitions lie in \(Q^{j,k}_{x,y}\). When \(j=3\) the pair \((x,y)\) can always be chosen with \(y\ne s(x)\): if \(|B\cap B'|=1\), there are two choices of \(y\), and if \(|B\cap B'|=2\), there are two choices of \(x\) with the same \(y\), and \(s\) is injective. The subfamily has \(1,5,21,49,116\) cliques for \(n=4,\dots,8\); for \(n\le7\) these are the minimum.

\textbf{The lower bound at \(n=8\).} The edge clique cover number is at least the optimum of its linear relaxation, the fractional edge clique cover \[\min\Bigl\{\sum_C z_C:\ \sum_{C\ni e}z_C\ge1\text{ for every edge }e,\ z\ge0\Bigr\},\] over the maximal cliques \(C\) of \(G\). Relabelling the leaves maps \(G\) to itself, so the average of an optimal solution over all relabellings is again optimal, and it gives every edge in a class of crossing pairs that relabelling maps to each other the same coverage. The optimum is therefore that of a linear program with one constraint per class: with \(N_o\) pairs in class \(o\) and \(c_o(C)\) of them inside \(C\), it is \(\min\{\sum_v t_v:\sum_v c_o(v)\,t_v\ge N_o\text{ for every }o,\ t\ge0\}\), over the distinct vectors \(v=(c_o(C))_o\). The class of a crossing pair is given by its four quadrant sizes, up to exchanging the two bipartitions and complementing either. At \(n=8\) there are \(8\) classes and \(12{,}160{,}639\) maximal cliques with
\(591\) distinct vectors; the optimum is \(784/9\approx87.1\), with optimal primal and dual solutions checked in exact arithmetic, so every edge clique cover has at least \(88\) cliques. The same computation gives \(5\), \(39/2\) and \(217/5\) for \(n=5,6,7\).

\section{Code and data availability}\label{code-and-data-availability}

The enumerator for Theorem 5, the compact solver, and the scripts that regenerate the tables of Section 8 are available at \url{https://github.com/Leonardini/BalancedMinEvolutionPaper}. The benchmark matrices are those distributed with the distance-indexed solver of Catanzaro \emph{et al.} {[}13{]}.

\section{Funding}\label{funding}

LC and SB acknowledge funding from the MRC Centre for Global Infectious Disease Analysis (reference MR/X020258/1), funded by the UK Medical Research Council (MRC). This UK funded award is carried out in the frame of the Global Health EDCTP3 Joint Undertaking.

\section{References}\label{references}

Each reference is marked by topic: \topicbme{} BME optimization and the BME polytope; \topicphylo{} phylogenetics and tree metrics; \topicpoly{} polyhedral combinatorics, optimization, information theory and numerical computation; \topicsoftware{} software. References are numbered in order of first citation.

\protect\phantomsection\label{refs}

\end{document}